\documentclass[12pt]{article}
\pdfoutput=1

\usepackage{color}
\usepackage{amsmath}
\usepackage{amsfonts}
\usepackage{amssymb}
\usepackage{caption}
\usepackage{graphicx}
\usepackage{slashed}            
\usepackage{subcaption}
\usepackage{xspace}				
\usepackage{cite}
\usepackage{bm}
\let\originalleft\left
\let\originalright\right
\renewcommand{\left}{\mathopen{}\mathclose\bgroup\originalleft}
\renewcommand{\right}{\aftergroup\egroup\originalright}

\usepackage[english]{babel}
\usepackage{fancyhdr}
\usepackage{amsmath}
\usepackage{amssymb}
\usepackage{amsfonts}
\usepackage{psfrag}
\usepackage[applemac]{inputenc}
\usepackage{graphicx}

\newcommand{\arXiv}[2]{\href{http://arxiv.org/pdf/#1}{{\tt #2/#1}}}
\newcommand{\arXivold}[1]{\href{http://arxiv.org/pdf/#1}{{\tt #1}}}

\renewcommand{\tilde}{\widetilde} 

\newcommand{\beq}{\begin{eqnarray}}
\newcommand{\eeq}{\end{eqnarray}}

\newcommand{\bag}{\begin{align}}
\newcommand{\eag}{\end{align}}

\usepackage[hypertexnames=false]{hyperref}

\begin{document}
\begin{titlepage}

\begin{center} 
{\huge \bf Self-Organized Higgs Criticality} 
\end{center}
\vspace*{0.4cm} 

\begin{center} 
{\bf \  Cem Er\"{o}ncel, Jay Hubisz, Gabriele Rigo} 

 {\it Department of Physics, Syracuse University, Syracuse, NY  13244}

\vspace*{0.1cm}
{\tt  
 \href{mailto:ceroncel@syr.edu}{ceroncel@syr.edu},
 \href{mailto:jhubisz@syr.edu}{jhubisz@syr.edu},  
 \href{mailto:garigo@syr.edu}{garigo@syr.edu}}

\end{center}

\vglue 0.3truecm

\begin{abstract}
The critical point for a Higgs sector can be a point of interest in the potential for a modulus field such as the radion of an extra dimensional construction, or the dilaton of spontaneously broken approximate conformal invariance.  In part motivated by conjectures about the self-organized critical state in statistical physics, we construct a 5D model in which there is an infrared emergent Higgs instability due to violation of the Breitenlohner-Freedman bound deep in the interior of a near AdS geometry.  This is holographically dual to a ``running" scaling dimension transitioning from real to complex with decreasing scale.  The complex scaling indicates an instability to be resolved by condensates which modify the interior geometry and backreact on the 5D radion potential.  Studying the model at small gravitational backreaction, we find a rich possible vacuum structure and uncover evidence that resolution of the instability requires a non-trivial cosmology.
\end{abstract}

\end{titlepage}

\setcounter{equation}{0}
\setcounter{footnote}{0}
\setcounter{section}{0}

\section{Introduction}
\label{sec:intro}

The Higgs instability in the electroweak sector of the Standard Model (SM) appears thus far to be of the simplest variety, with the Higgs sector residing ``unnaturally" close to a critical point seemingly unprotected by symmetry and well described by mean field theory.  The  Higgs sector is associated with a Landau-Ginzburg theory of a symmetry breaking phase transition.  The Higgs field of the Standard Model develops a vacuum expectation value (VEV) due to a relevant operator (the Higgs mass term) having the ``wrong sign" in the infrared which destabilizes the origin in field space.  The naturalness issue, or the hierarchy problem, is the statement that in units of the much larger fundamental scales in the problem, i.e.~the Planck or GUT scales, the bare mass must be tuned to an absurd degree to accommodate the observed value for the VEV and the mass of the Higgs particle, and that quantum corrections spread this sensitivity among fundamental parameters.  That is, quantum effects make the Higgs mass similarly sensitive to, for example, the value of the top quark Yukawa coupling.  As a consequence it is expected that low energy effective theories with Higgs sectors like that of the SM are extraordinarily rare when there are other large physical scales present.

Most proposed resolutions of this problem invoke new symmetries requiring new particles with fixed interactions that ameliorate the sensitivity of the Higgs mass to larger mass scales.  The paucity of new particles at the TeV scale has sown growing doubt that this is the way nature has created a low electroweak scale.  

In this work, we begin an investigation of the possibility that the critical point for Higgs sectors like that of the SM can arise naturally not due to symmetry, but rather because of a self-organization principle, as suggested in~\cite{Giudice:2008bi}.  This is inspired by critical behavior that has been observed in a wide variety of seemingly unrelated physical systems in nature~\cite{soc,socflicker}, with the canonical example being the sandpile.  In this example, a pile of sand is created and sustained by slowly pouring sand from a funnel onto a fixed point.  The sand self-organizes into a cone with opening angle fixed by microscopic interactions between grains.  Perturbations, even if only involving a single grain of sand, result in avalanches at all scales, following a power law distribution.   After the avalanche, when the system has removed the perturbation,  the cone of sand is in a new configuration, but still critical so long as sand continues to be slowly added.  Such ``self-tuning" is also thought to arise at earthquake fault lines, river bifurcations, and temporally near financial market crashes~\cite{25YearsofSOC}.  It has been key to describing some astrophysical phenomena as well~\cite{Aschwanden:2014dna}.  It has been hypothesized that these phenomena of criticality at the threshold of ``catastrophic failure" can be related to an instability arising from critical exponents becoming complex, corresponding to discrete scale invariance~\cite{Sornette:1997pb}.  This possibly suggests an approach to the Higgs fine-tuning problem in extra-dimensional models in AdS space, where similar features appear when the Breitenlohner-Freedman (BF) bound~\cite{Breitenlohner:1982jf} is violated, leading to complex scaling dimensions and a complete loss of conformality in a hypothetical 4D dual~\cite{Kaplan:2009kr}.

Typically, in the examples we have in statistical physics, criticality is ``self-organized" by temporal loading of the system:  the slow addition of sand grains at a pile's apex, gradual stress building at faults due to tectonic drift, etc.  In relativistic theories, this slow temporal loading can be exchanged with mild spatial gradients, and in 5D, through the AdS/CFT duality~\cite{Maldacena:1997re}, spatial translation and gradients can be, in turn, related to scale transformations and renormalization group evolution.  This further suggests an approach where the instability associated with complex scaling dimensions is reached dynamically through slow renormalization group evolution, or, in a 5D dual, via growth of a deformation of AdS space that leads to eventual violation of the BF bound.  This has a parallel in some perturbative 4-dimensional theories where dimensional transmutation directs the Higgs potential~\cite{Coleman:1973jx}. In models such as the MSSM, the instability of the electroweak sector can be arrived at through quantum corrections~\cite{Ibanez:1982fr,Ellis:1983bp,AlvarezGaume:1983gj}:  RG flow from some microscopic theory seemingly devoid of instabilities evolves to an IR theory where a Higgs picks a non-vanishing condensate.

Stepping away from the background motivation, we can state generally that the eventual goal of this line of research is to have a zero for the Higgs mass term coinciding (or nearly coinciding) with a minimum in the potential for an extra-dimensional modulus field over a wide range of extra-dimensional input parameters.  In this case Higgs criticality would be a dynamical attractor for the theory without extreme sensitivity to fundamental constants.  In other words, such a 5D system \emph{self-tunes} to the critical point of its 4D low energy effective Higgs theory.  This is similar to the way in which the strong CP problem is resolved by the axion hypothesis where the axion field promotes the CP phase to a dynamical field whose potential minimum resides at the point where CP is conserved in QCD, and also to models which solve the hierarchy problem using cosmological dynamics and slightly broken shift symmetries~\cite{Graham:2015cka}.

\begin{figure}
	\centering
	\includegraphics[width=.75\textwidth]{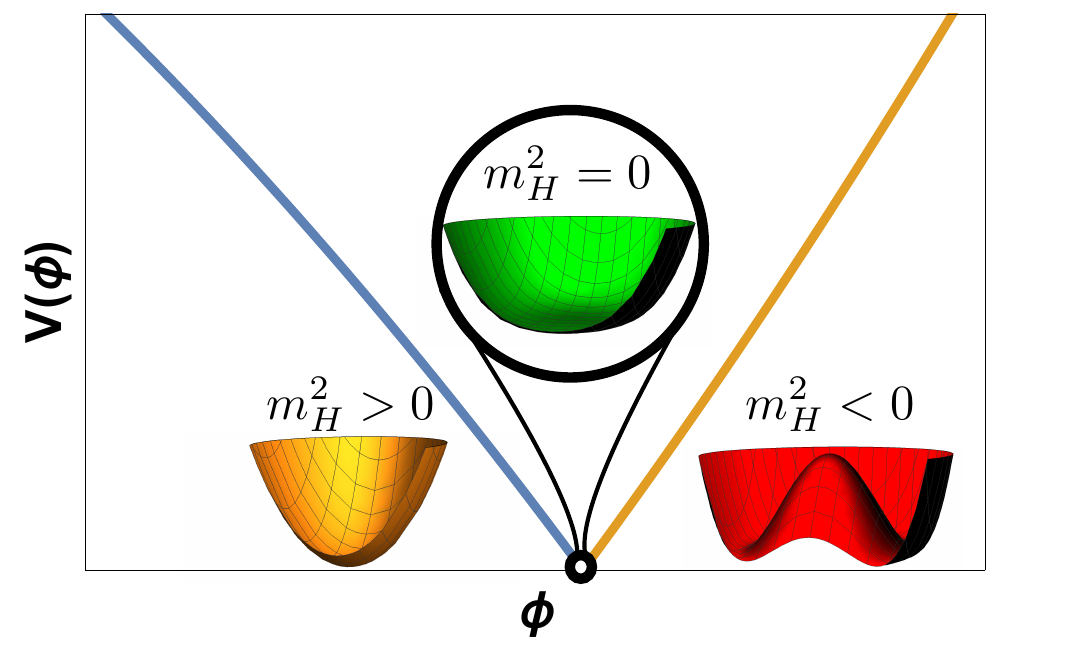}
	\caption{\small This figure exhibits a cartoon of a potential for a modulus field, $\phi$, where the singular minimum matches on to a critical point at which the mass of a physical light Higgs field fluctuation passes through zero.  On either side of the singular point, the Higgs boson mass is finite and positive, but on one side the mass squared for the field is negative, with the instability driving spontaneous symmetry breaking.}
	\label{fig:singularhiggs}
\end{figure}

To be properly identified as self-organization, in the most conservative application of this label, criticality near the minimum of the potential must be robust under reasonable variation of model input parameters: we are looking for a theory with a near-critical Higgs \emph{region}.  From the perspective of the low energy theory, at energy scales below the modulus VEV, the lightness of the Higgs would be in apparent violation of effective field theory principles and power counting, that is, a violation of the principle of locality of scales.\footnote{Such violation has been observed before in a toy model with discrete scale invariance~\cite{Georgi:2016tjs}, another suggestive connection to self-organized critical systems.}  This could be reconciled by the fact that in a microscopic 5D theory, quantum corrections might renormalize the potential for the Higgs and modulus field simultaneously,  shifting the \emph{location} of the minimum, but not the property that the physical Higgs boson is light at that minimum.  A cartoon is shown in Figure~\ref{fig:singularhiggs}.  We emphasize that this has not yet been realized in a dynamical model in this work, but rather is a longer term goal in which this paper could be a crucial first step.

To realize such a scenario, there first must be a feedback mechanism, such that when the Higgs field develops a VEV, the potential for the modulus responds, as shown in Figure~\ref{fig:singularhiggs}.  This is somewhat akin to the way thresholds can play an important role in Casimir energies, as in, for example~\cite{ArkaniHamed:2007gg}.  If this feedback is itself the origin of the minimum of the potential, with the minimum being at or near the critical value at which the Higgs VEV turns on, the modulus field will be attracted to the minimum, where the mass of the Higgs fluctuation is small or vanishing.  


In one ``toy" and one fully dynamical construction, we investigate the mutual potential of a modulus field (in our case, the radion of a Randall-Sundrum (RS) 2-brane model~\cite{Randall:1999ee}) and a 5D Higgs field.  The models are constructed so that the value of the modulus field vacuum expectation value influences the Higgs stability criterion.  Specifically, we consider scenarios where the Higgs instability is linked to dynamical violation of the BF bound  far from the UV boundary of the RS geometry through interactions with the modulus field.

The CFT dual of this picture corresponds to the theory described above:  one which is conformal in the deep UV, without instabilities, and which contains a near marginal deformation that drives a slow running of the scaling dimension of an operator in this approximate CFT~\cite{Rattazzi:2000hs,ArkaniHamed:2000ds}.  There is an interplay between two operators, ${\mathcal O}_\epsilon$, a near marginal operator, and ${\mathcal O}_H$.  Similar to the way dynamical symmetry breaking for the Higgs can occur via radiative corrections, here an analogous instability is reached in the RG flow for the dual picture, with the scaling dimension for the operator ${\mathcal O}_H$ becoming complex, and causing an ``unhealthy"~\cite{Luty:2012ww} limit cycle-like behavior in the RG flow that is terminated by condensation of operators in the approximate CFT, breaking the approximate conformal invariance spontaneously.  

Curiously, in the models studied here, a nontrivial cosmology may be a common feature of the classical ground state.  In the 5D picture, the metric and scalar fields must include a cosmology for the low energy 4D effective theory in order to satisfy the metric junction conditions for the two brane model, unless UV and IR brane tensions are tuned.  We suspect this is due to a frustration mechanism similar to that which causes striped and other inhomogeneous phases to appear in some high $T_c$ superconductors~\cite{stripes}, and may lead in this case to a time-dependent vacuum state~\cite{Shapere:2012nq,Wilczek:2012jt}.  Studies of holographic superconductivity have identified such phases when the BF bound is violated deep in the interior of asymptotically AdS geometries due to non-trivial gauge and gravitational field backgrounds~\cite{Hartnoll:2016apf}.  In other words, the resolution of the RG limit cycle instability may have its imprint in the low energy theory as a limit cycle in time evolution. We have left a full exploration of the time dependence for future work, however we do speculate on the possible relevance of this feature for cosmology~\cite{Bains:2015gpv,Easson:2018qgr}.

The article is organized as follows:  In Section~\ref{sec:eft}, we set the stage with a discussion of a possible effective theory for a dilaton vacuum expectation value that has features that we find in our holographic model, and which points towards a new way to set the confinement scale in an approximately conformal theory.  In Section~\ref{sec:runningmass}, we describe a 5D  toy holographic implementation of these ideas that leaves out certain backreaction terms, with a bulk Higgs mass that is given \emph{explicit} dependence on the extra-dimensional coordinate.  In Section~\ref{sec:gwdriven}, we show how to promote this simplest model to a  dynamical one, where the effective Higgs mass evolves due to a coupling to a ``driving" scalar which picks a coordinate-dependent VEV.  In Section~\ref{sec:CFTInterpretation} we discuss in more detail a possible approximate CFT interpretation of the 5D model, and make some connections to generalized BKT scaling discussed in~\cite{Kaplan:2009kr}.  In Section~\ref{sec:discussion}, we briefly discuss connections to self-organized critical models in statistical physics, to catastrophic failure modes of these systems, and future implementation of these ideas into an extension of the SM.  Finally, we discuss aspects of the cosmology of the model, part of which involve speculation that can be more fully resolved with future work, in Section~\ref{sec:cosmo}.

\section{Preliminaries:  The Frustrated Dilaton}
\label{sec:eft}
We begin by considering an effective theory of a dilaton that has aspects of the behavior of the more detailed holographic models we consider.  In an approximately scale-invariant theory that undergoes spontaneous breaking, the potential that sets the scale of that breaking is a (nearly) scale-invariant quartic:  $V_\text{dil} = \alpha(f) f^4$, where $\alpha$ has some weak dependence on $f$ that encodes the amount of explicit breaking.  This potential can have a nontrivial minimum at small values of $f$ if $\alpha$ obeys some basic properties.

It is important to note that in this effective theory, $f$ is a summed total scale of conformal breaking, and relations between individual VEVs are hidden in this potential.  In other words, many operators could be simultaneously condensing which all contribute to conformal breaking.  In addition, if there is some small breaking of conformal invariance, there can be nontrivial relations between these operator VEVs that are not encoded in the total dilaton potential.

Now we note that it could be the case that there is simply no solution that satisfies the relation between VEVs in the approximate CFT for a total $f$ below some critical value, $f < f_\text{crit}$.  How does the theory exit the CFT in this case?  The dilaton is frustrated -- internally, it seeks to satisfy the relations between operator VEVs, providing a lower limit to the total breaking scale.  On the other hand, in the low energy theory, a smaller or vanishing $f$ is preferred.  

As a very simple example of this, we could consider two VEVs $f_\phi$ and $f_\text{higgs}$ where at some critical value of decreasing $f_\phi$, $f_\text{higgs}$ begins to turn on:
\beq
f_\text{higgs}^2 = \left\{ \begin{array}{ll} 
	-\frac{\lambda f_\phi -\mu^2}{\lambda_H} & \lambda f_\phi - \mu^2 < 0 \nonumber \\
	0 & \text{otherwise}
\end{array}. \right.
\eeq
Such behavior occurs in the ``relaxion" models of electroweak symmetry breaking.  Note however, that we are not specifying a global potential for this behavior, but instead are merely providing an ad hoc relationship between operator VEVs that might arise inside of the approximate CFT.  In the holographic models we consider, $f_\phi$ corresponds to the VEV of a marginally relevant operator, and $f_\text{higgs}$ to the VEV of an operator whose scaling dimension is driven into the complex plane.

If these VEVs contribute to an effective dilaton potential as $V_\text{dil} \approx \alpha (f_\phi^4+f_\text{higgs}^4)$ (as though they are both from operators with scaling dimension near 4), the potential will be globally minimized when $f_\phi^2 = \frac{\lambda}{2 \lambda_H f_\phi} f_\text{higgs}^2$ if the relationship between VEVs is maintained.  A large hierarchy between $f_\text{higgs}$ and $f_\phi$ can be created by having a small value of $\lambda_H$, and the dilaton potential appears to be minimized at some $f_\text{crit}$ while enforcing the relation.

 We note that the dilaton potential itself does not enforce the relation between the VEVs, which is instead specified by some dynamics that is part of the UV completion of the dilaton effective theory.  This can create a puzzle from the perspective of the low energy theorist, which sees only the total breaking scale $f$, not the interrelations between the individual $f$'s that contribute.  The low energy observer would think that $f = 0$ should minimize the pure quartic potential, but instead the theory gets ``trapped" at a larger $f$.  Of course these relations between VEVs would require some breaking of scale invariance, and the potential would not be precisely quartics.  

There may be a connection here to the concept of frustrated phase separation in condensed matter physics, a phenomenon where (for example) high $T_\textnormal{c}$ superconductivity is blocked due to long range Coulomb interactions, and the theory resolves the tension by creating an intermediate phase which spontaneously breaks translation invariance, creating ``stripes"~\cite{stripes}.  Here, Higgs condensation may be blocked by quasi-long range dilaton interactions, and there may be some analog of these striped phases that resolves the frustration that is important for cosmology.

With this hypothetical scenario as context and motivation, we begin our holographic studies.
\section{Toy Model:  Explicitly Varying Higgs Mass}
\setcounter{equation}{0}
\label{sec:runningmass}
To illustrate a basic model with the features we seek, we work in 5D asymptotically anti-de Sitter space without a UV brane, and give the Higgs a bulk mass term that varies \emph{explicitly} with the extra-dimensional coordinate.  We strongly emphasize that this is a toy model that easily illustrates some curious features that help motivate the more realistic model in Section~\ref{sec:gwdriven}.  Specifically, this model leaves out backreaction between the Higgs and the (here unspecified) dynamics that give rise to the varying mass term, while in Section~\ref{sec:gwdriven}, this backreaction is taken fully into account.  

The metric can be written as (setting the AdS curvature near the AdS boundary, $k$, to 1)~\cite{Bunk:2017fic}:
\beq
ds^2 =  \frac{1}{z^2} \left[ dx_4^2 - \frac{dz^2}{G(z)} \right].
\label{eq:metansatz}
\eeq
The coordinate $z$ ranges from $0$ at the AdS boundary to an IR brane at $z=z_1$, and for $z \rightarrow 0$, the function $G$ has the asymptotic behavior $G(z) \rightarrow 1$ and $G'(z) \rightarrow 0$.  Away from $z = 0$, the function $G$ encodes the effects of gravitational backreaction due to nontrivial bulk physics such as condensates.  We are restricting our ansatz for the background to solutions obeying 4D Lorentz invariance.

The action is given by
\beq
\begin{aligned}
	S &= \int d^4 x dz \sqrt{g} \left[  |\partial_M H|^2 +\frac{6}{\kappa^2}-m^2(z)|H|^2 -\frac{1}{2 \kappa^2} R \right] \\
	&\phantom{{}={}}- \int d^4 x z^{-4} m_0^2 |H|^2 \bigg\rvert_{z\rightarrow 0}-\int d^4 x z^{-4} V_1 (|H|) \bigg\rvert_{z\rightarrow z_1},
\end{aligned}
\eeq
with $\kappa^2 = 1/(2 M_\text{Pl}^3)$.  The bulk mass function is chosen to a have fixed value in the limit $z \rightarrow 0$, and decreases monotonically and slowly as $z$ increases:\footnote{It is not difficult to arrange for this type of $z$-dependent mass term to arise dynamically, rather than through this forced explicit breaking of the isometries of AdS.  We give examples in Section~\ref{sec:gwdriven}.}  
\beq
m^2(z) = -4 + \delta m^2 - \lambda z^\epsilon.
\eeq
Note that $m^2 = -4$ corresponds to the Breitenlohner-Freedman bound, and $\delta m^2$ is taken to be a positive quantity so that the $z \rightarrow 0$ limit is well-defined.\footnote{If the mass is taken below the BF bound as $z \rightarrow 0$, perturbations solving the scalar equation of motion oscillate rapidly in the UV, indicating the need for an ultraviolet cutoff, such as a brane that cuts off the small $z$ region of the spacetime~\cite{Kaplan:2009kr}.}  We note that past work explored constant Higgs bulk mass at or near the BF bound~\cite{Vecchi:2010em,Geller:2013cfa} with interesting implications for radius stabilization.  A possible relationship between the BF bound and scalars with suppressed mass in lattice studies of theories at the boundary of the conformal window was discussed in~\cite{pomarolplanck}.  For the IR brane potential, we take 
\beq
V_1 (|H|) = T_1+\lambda_H |H|^2 \left(|H|^2 - v_H^2\right),
\eeq
where $T_1$ is the tension of the brane.  
The Higgs may, for some regions of parameter space, pick a nonvanishing vacuum expectation value, $\langle H \rangle = \phi(z)/\sqrt{2}$, where the Higgs VEV has a nontrivial profile along the $z$-coordinate.

Restricting to solutions that obey 4D Lorentz invariance, the 5-5 component of the Einstein equations relate the metric function $G$ to the behavior of the Higgs VEV in the bulk:
\beq
G = \frac{ -\displaystyle\frac{\kappa^2}{6} V(\phi)}{1- \displaystyle\frac{\kappa^2}{12} (z \phi')^2} ,
\eeq
and they can be further employed to reduce the effective potential for the classical background configuration to a pure IR boundary term~\cite{Bellazzini:2013fga}:
\beq
V_\text{rad} = \frac{1}{z_1^4} \left[ V_1 (\phi) + \frac{6}{\kappa^2} \sqrt{G} \right].
\eeq
There is generally a UV contribution from the brane at $z_0$ as well, but it vanishes as $z_0^{2 \sqrt{\delta m^2}}$ in the $z_0 \rightarrow 0$ limit, with the exception of a constant term which is tuned to give vanishing effective cosmological constant.  We also note that the remaining components of Einstein's equations in the bulk do not give additional information on $G$, being equivalent to the scalar field equation of motion for the $z$-dependent background, and that for the purposes of calculation of the effective potential for the size of the extra dimension, we fix the 4D portion of the metric to be flat.  This amounts to satisfying vanishing variation of the action with respect to those metric components in a trivial manner, requiring that the variations themselves vanish:  $\delta g_{\mu\nu} = 0$.

For small values of the Higgs VEV, or alternatively weak 5D gravity, this effective radion potential reduces to
\beq
V_\text{rad} = \frac{1}{z_1^4} \left[ V_1 (\phi) + \frac{6}{\kappa^2} - \frac{1}{4} m^2 (z_1) \phi^2(z_1)+ \frac{1}{4} z_1^2 \phi'^2(z_1) \right].
\label{eq:radpot}
\eeq
The scalar field equation of motion in the limit of small $\kappa^2$ is
\beq
\phi'' -  \frac{3}{z} \phi'-\frac{1}{z^2} \frac{\partial V}{\partial \phi} =0,
\eeq
with energetic favorability of a nontrivial solution depending on the boundary conditions, which are (again in the small $\kappa^2$ limit):
\beq
z\phi' |_{z = z_{0,1}} = \pm \frac{1}{2} \frac{\partial V_{0,1}(\phi)}{\partial \phi}.
\eeq
Near the AdS boundary $z\rightarrow 0$, the solutions are power law in $z$, with the expected behavior $\phi \propto z^{2 \pm \sqrt{\delta m^2}}$, where the two different scaling laws correspond to two different boundary conditions or definitions of the action~\cite{Klebanov:1999tb}.  The scaling law $z^{\Delta_+}$ is more generic, with fine-tuning of BCs (or supersymmetry) required to obtain the scaling with power law $\Delta_-$.  The full solution for all $z$, including the effects of the changing bulk mass, is
\beq
\phi \sim \phi_\pm z^2  J_{\pm \nu} \left( \frac{2 \sqrt{\lambda}}{\epsilon} z^{\epsilon/2} \right),
\label{eq:hvevsol}
\eeq
where $\nu\equiv2 \sqrt{\delta m^2}/\epsilon$. For $m_0^2 \ne 2-\sqrt{\delta m^2}$, only the $+$ solution is relevant, while for the special case $m_0^2 = 2-\sqrt{\delta m^2}$, the field behavior is given by the $-$ solution.  Choosing the special case corresponds in the holographic picture to fine-tuning the coefficient of an operator $O_H^\dagger O_H$ to force the RG flow to go ``backwards" compared to the more generic $+$ scaling solution~\cite{Kaplan:2009kr,Gubser:2002vv}, or in other words, tuning so that the theory sits at a UV fixed point.

Choice of the UV boundary condition does not much affect the discussion, and we choose to display the effects of the first of these two solutions, taking $m_0^2 = 0$.

For large values of $z$, and small $\epsilon$, the asymptotics of the Bessel function exhibit log-periodicity on top of scaling when $z$ is past the point where the BF bound is surpassed by the evolving bulk mass:
\beq
\phi \propto z^{2-\epsilon/4} \cos \left( \sqrt{\lambda} \log z + \gamma \right).
\eeq
This log-periodic power law behavior may be a common feature in systems where criticality is self-organized~\cite{Sornette:1997pb}.  

The condition for formation of a condensate is met when the IR brane boundary condition favors a nontrivial value for the coefficients of the bulk solution:
\beq
\frac{1}{\epsilon} \left( \lambda_H v_H^2 - 4 \right) \ge \frac{x J'_\nu (x)}{J_\nu(x)},
\eeq
where $\nu \equiv 2 \sqrt{\delta m^2}/\epsilon$ and $x \equiv 2 \sqrt{\lambda} z_1^{\epsilon/2}/\epsilon$.  Equality is associated with the presence of the exact critical point.  Note that equality is satisfied at many values of $z_1$ due to quasi-periodicity of the right-hand side at large $z_1$.  We label the $i$-th critical point as $z_\textnormal{c}^i$.  The emergence of a massless degree of freedom at these critical points can be seen in the small momentum behavior of the bulk correlator for the Higgs fluctuations.  Working in the unbroken phase, the Green's equation for scalar fluctuations is given by
\beq
\left[ \partial_z^2 +p^2 - \frac{3}{z} \partial_z - \frac{m^2(z)}{z^2} \right] G\left(z,z';p^2\right) = i z \delta(z-z').
\eeq
At the point of criticality, for small $p^2$, the Green's function in terms of an eigenfunction decomposition takes the form
\beq
G\left(z,z';p^2\right) \approx \frac{\psi_0(z) \psi^*_0(z')}{p^2} - \sum_{i=1}^{\infty} \frac{\psi_i(z) \psi^*_i (z')}{m_n^2},
\eeq
where $\psi_0(z)$ solves the 5D equation of motion for $p^2 = 0$, and thus takes the same functional form as the VEV, $\phi(z)$.  The $m_n$ are the usual KK-mode masses.  As we discuss in further detail below, the $m_n^2$ are not guaranteed to be positive at all of the critical points, and in fact only one, the smallest $z_\textnormal{c}$ critical point, $z_\textnormal{c}^1$ has a positive spectrum.

As the Higgs VEV turns on, the character of the radion potential also changes dramatically.  Of primary interest is the behavior of the radion potential near the region of $z_1$ where the Higgs VEV is just turning on.  A linearized approximation of the Higgs contribution to the potential gives the leading contribution in the immediate neighborhood of criticality.  The function for $\phi(z_1)^2$ is analytic, and near its zeros, say one at $z_1=z_\textnormal{c}^1$, we have
\beq
\phi (z_1)^2 \approx \sigma^2 \left(\frac{z_1}{z_\textnormal{c}^1}-1\right),
\eeq
where $\sigma^2$ is a positive function of the parameters of the theory:
\beq
\sigma^2 = \frac{-4 m^2 \left(z_\textnormal{c}^1\right) + \lambda_H v_H^2 \left(\lambda_H v_H^2- 8\right)}{2 \lambda_H}.
\eeq
The radion potential in the regime just after the VEV turns on is given by
\beq
V_\text{rad} \approx \frac{1}{z_1^4} \left[ \delta T_1 + \frac{\lambda_H\sigma^4}{8} \left(\frac{z_1}{z_\textnormal{c}^1}-1\right) \right].
\eeq
The radion potential is thus piecewise, and if we look in the vicinity of the critical point, it takes the form
\beq
V_\text{rad} \approx \left\{ \begin{array}{ll}
	\displaystyle\frac{1}{z_1^4} \delta T_1 & z_1 < z_\textnormal{c}^1 \\
	\displaystyle\frac{1}{z_1^4} \left[ \delta T_1 + \frac{\lambda_H\sigma^4}{8} \left(\frac{z_1}{z_\textnormal{c}^1}-1\right) \right] & z_1 > z_\textnormal{c}^1 \end{array} \right.,
\eeq
where $\delta T_1$ is the mistune between the IR brane tension and the bulk cosmological constant.  We note that quantum corrections will spread $z$-dependence from the scalar mass to the cosmological constant term, and the background will no longer be pure AdS.  These small corrections will modify the background potential away from a pure quartic - this is not, however, important for the features we draw attention to in this model.  In the next section, the background potential is modified away from a pure quartic by a Golderberger-Wise type potential.  

\begin{figure}[t]
	\centering
	\begin{subfigure}[t]{0.49\textwidth}
		\includegraphics[height=2.0in]{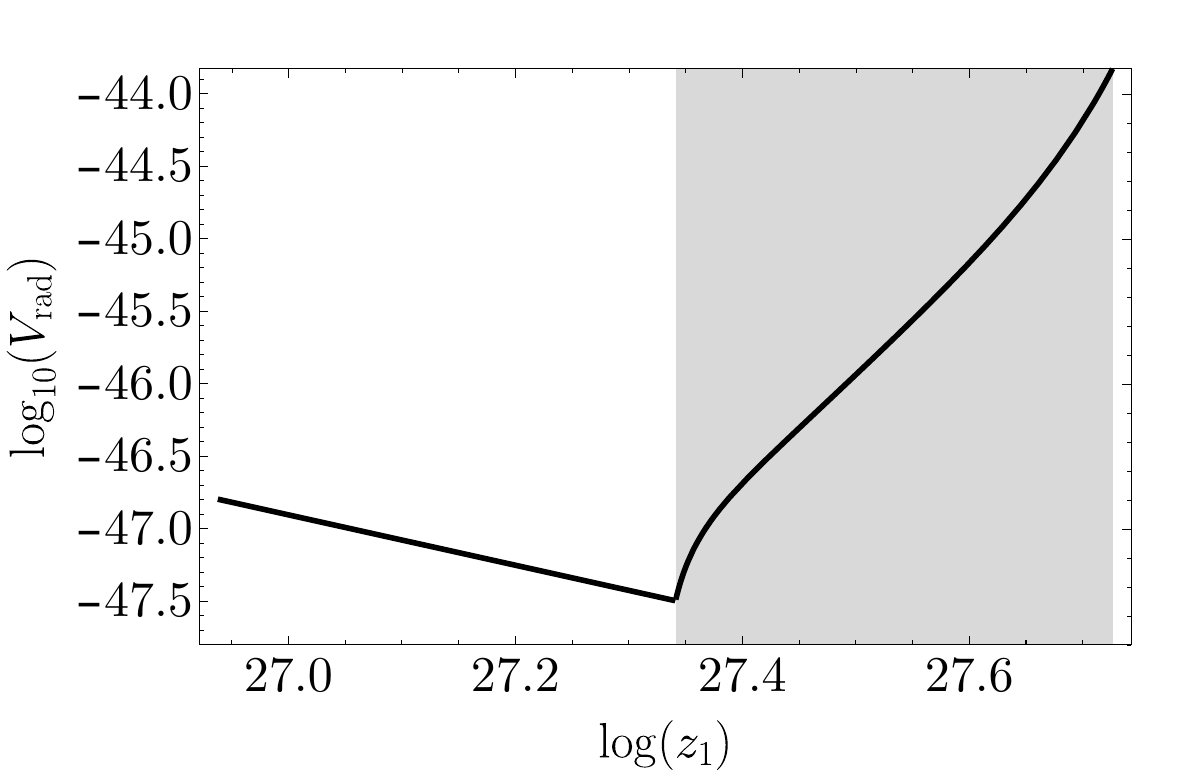}
	\end{subfigure}
	\begin{subfigure}[t]{0.49\textwidth}
		\includegraphics[height=2.0in]{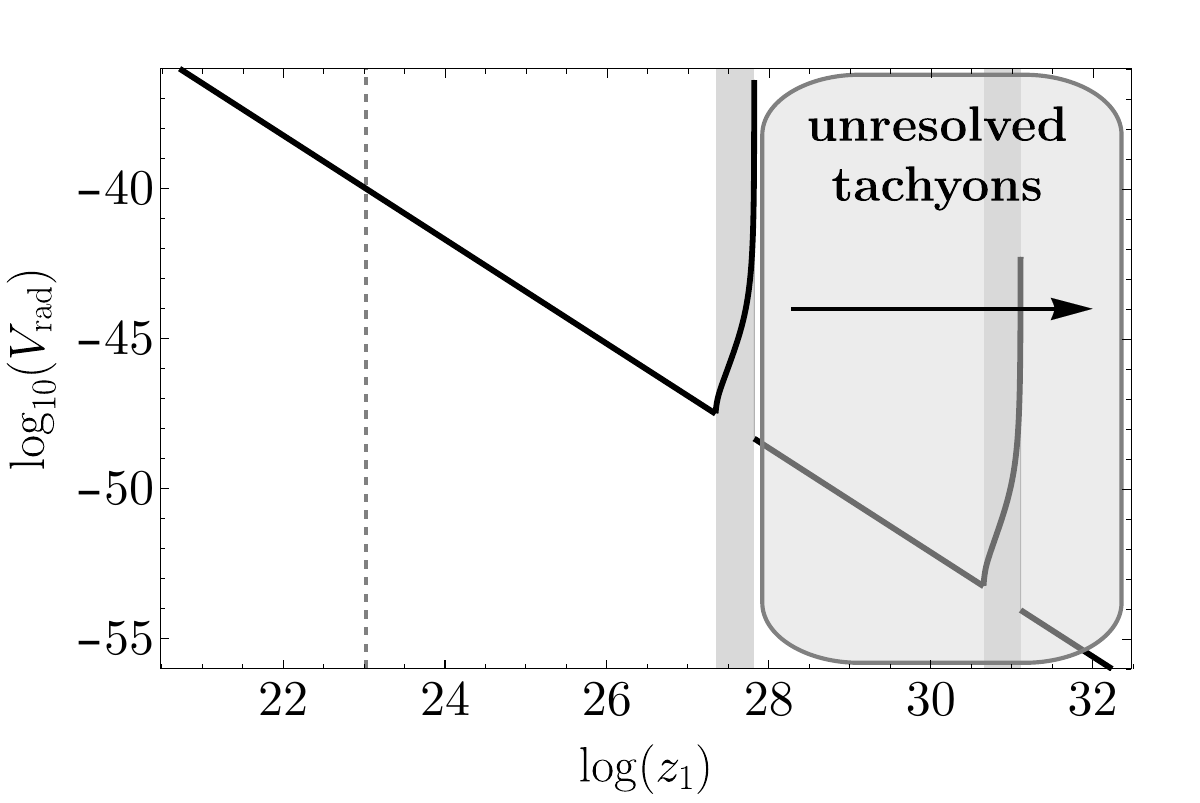}
	\end{subfigure}
	\caption{\small Here we display the radion potential, $V_\text{rad}(z_1)$.  In the white region, the Higgs VEV is vanishing, and the radion potential is a pure quartic.   In the gray region, $\phi(z_1) \ne 0$, and the contribution of the Higgs to the radion potential causes a kink-like minimum to appear at the critical points.  In the first plot, we have zoomed in on the first minimum, corresponding to the smallest $z_1$ for which the criticality conditions are met.  In the second plot, we zoom out, showing other potential minima.  These are unhealthy, in that the theory at this point contains unresolved tachyons.  The dashed vertical line in the second plot corresponds to the value of $z$ at which the evolving bulk Higgs mass passes the BF bound.}
	\label{fig:dilpot}
\end{figure}

 At the critical point, there is a kink discontinuity in the radion potential.\footnote{We note that this kink feature is due to the fact that we are deliberately ignoring backreaction between the Higgs and the dynamics that creates the varying mass term.  The properties of the critical point differ in complete models, and we begin a study of an explicit example in Section~\ref{sec:gwdriven}.}
The contribution of the Higgs condensate to the radion potential is also positive definite.  In order for the derivative of the radion potential to change sign, creating a kink \emph{minimum}, an additional condition for the radion quartic (the IR brane mistune) must hold:
\beq
0 < \delta T_1 < \frac{1}{128 \lambda_H} \left[ 4 m^2 \left(z_\textnormal{c}^1\right) - \lambda_H v_H^2 \left( \lambda_H v_H^2 -8 \right) \right]^2.
\eeq
These requirements are satisfied relatively robustly under variation of the input parameters.  In Figure~\ref{fig:dilpot}, we show an example of the radion potential, where we have taken $\delta T_1 = 1$,  $\delta m^2 =1$, $\lambda = \epsilon = 0.1$, $v_H^2 = 0$, and $\lambda_H = 1/8$.

There are two plots in the figure: in the first, on the left, we show a close-up that focuses on the critical value of the radion VEV.  On the right, we zoom out.  Unsurprisingly, it appears that there are multiple such minima, and that the first one is metastable.  This is due to the quasi-periodicity of the Higgs VEV solution at large values of $z$.  Closer examination of the theory in these regions shows that there are unresolved tachyon instabilities associated with Higgs fluctuations.  In these regions, no VEVs are formed, but there are solutions to the scalar equation of motion with negative mass squared, as we show in the next subsection.

\begin{figure}[t]
	\centering
	\begin{subfigure}[t]{0.49\textwidth}
		\includegraphics[height=2.00in]{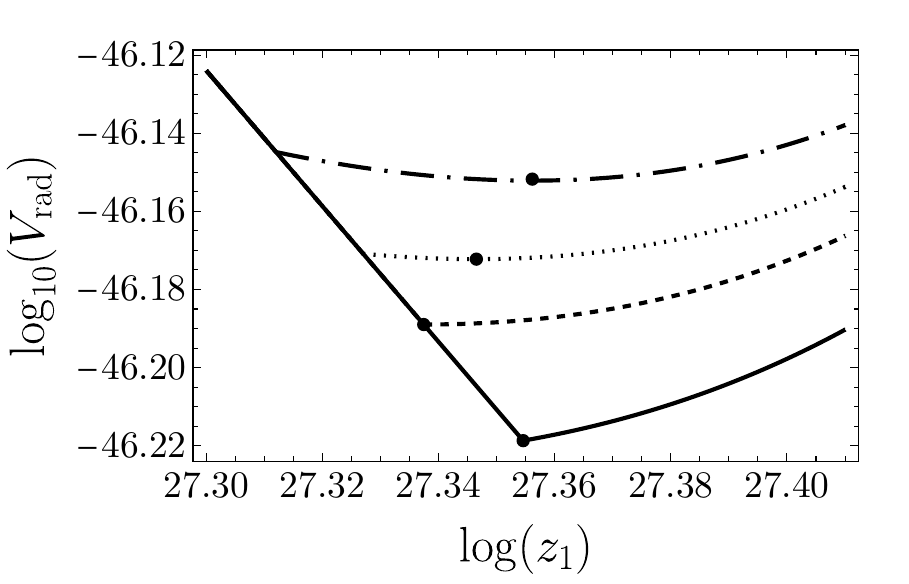}
	\end{subfigure}~
	\begin{subfigure}[t]{0.49\textwidth}
		\includegraphics[height=2.00in]{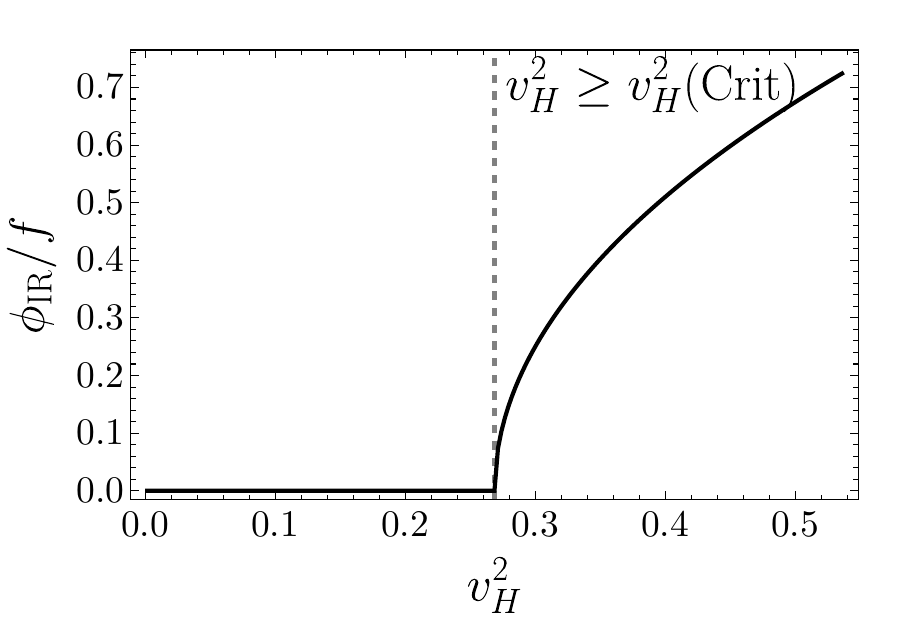}
	\end{subfigure}
	\caption{\small Here we show, on the left, the dependence of the potential on the IR brane parameter $v_H^2$ in the vicinity of the first critical point.  The curves correspond to $v_H^2 = -1$ (solid), $v_H^2 = v_H^2 (\text{crit})$ (dashed), $v_H^2 = 1$ (dotted), and $v_H^2 = 2$ (dot-dashed).  The dots indicate the minimum of the potential.  The minimum moves into the region where the Higgs VEV is nonzero after some critical point $v_H^2 (\text{crit})$.  On the right, we show the value of the Higgs field on the IR brane in units of the scale $f=z_\text{min}^{-1}$, where $z_\text{min}$ is the location of the minimum of the radion potential.  The VEV (and Higgs mass/inverse of the correlation length), which is proportional to $\phi_{\text{IR}}$, is vanishing below $v_H^2 (\text{crit})$, and grows quickly after the critical point is exceeded.}
	\label{fig:v2plots}
\end{figure}

Before studying the instabilities, it is worthwhile to explore the behavior of the effective potential under variations of the fundamental parameters.  Crucial to the success of the model as one of self-organized Higgs criticality is the existence of a broad critical region, over which the Higgs remains light.  In Figure~\ref{fig:v2plots}, we examine the behavior of the radion potential under variations of the IR brane Higgs mass squared, encoded in $v_H^2$.  We see that with changing $v_H^2$, the location of the minimum is relatively constant, however there is a crucial value of $v_H^2$ past which the location of the minimum moves away from the kink in the potential.  The region of parameter space where the minimum resides at the kink is a critical region, as there is a zero mode Higgs for all values of $v_H^2 < v_H^2 (\text{crit})$.

In the next subsection, we comment on properties of this kink minimum in the effective radion potential, in particular those related to the metric ansatz that we have enforced.

\subsection{Metric Boundary Conditions}
In calculating the radion potential in Eq.~\eqref{eq:radpot}, we have imposed the boundary conditions for the scalar fields, but we have neglected the metric junction conditions on the branes, which enforce
\beq
\sqrt{G(z_0)} = \frac{\kappa^2}{6} V_0, \qquad \qquad \sqrt{G(z_1)} = -\frac{\kappa^2}{6} V_1.
\label{eq:junction}
\eeq
This is equivalent to having the UV brane and IR brane contributions to the effective potential separately vanish~\cite{Csaki:1999mp,Csaki:2000zn}:
\beq
\begin{aligned}
	\tilde{V}_\text{UV} &= V_0 - \frac{6}{\kappa^2} \sqrt{G_0} = 0, \\
	\tilde{V}_\text{IR} &= V_1 + \frac{6}{\kappa^2} \sqrt{G_1} = 0,
	\label{eq:vtildes}
\end{aligned}
\eeq
where we have defined $V_\text{rad} = \frac{1}{z_0^4} \tilde{V}_\text{UV}+\frac{1}{z_1^4} \tilde{V}_\text{IR}$.
In fact, these conditions are not satisfied for any values of $z$ in the above potential.  These conditions should be interpreted as consistency conditions for our metric ansatz, in which we have forced the metric to exhibit 4D Lorentz invariance so that we can interpret the result as a Lorentz-invariant 4D effective potential for the modulus field.  In terms of variation of the scalar-Einstein-Hilbert action, we have satisfied vanishing variation of the action by keeping the variations of the 4D metric components $ \delta g_{\mu\nu}$ themselves to be zero, and in so doing, Eq.~(\ref{eq:junction}) is no longer a constraint on the solution.

In the usual Goldberger-Wise scenario, the second of the two conditions in Eq.~(\ref{eq:junction}) is met automatically at the minimum of the potential, with the value of the size of the extra dimension being set by its solution.  The first is then arranged for by tuning (equivalent to the usual tuning of the bare cosmological constant to small values).  From this, we can roughly interpret the time dependence away from the minimum of the usual Goldberger-Wise potential as a combination of cosmological acceleration and oscillations of the stabilized radion.  The situation is quite different in the case of the potential described above.  At the kink minimum generated by the Higgs contribution, these junction conditions cannot both be met unless two tunings are performed -- both the bare cosmological constant and the mistune in the IR brane tension.  

This doesn't necessarily mean that the region where the kink is a minimum is forbidden, but rather tells us that once inside this region, a dynamical geometry is unavoidable, and must be included in a fully consistent calculation of the spectrum of the low energy theory.  In other words, in a theory with fully dynamical gravity, the ansatz for the background \emph{must} be relaxed to include a nontrivial 4D cosmology.   In Section~\ref{sec:cosmo}, we discuss this issue of a dynamic cosmology further, and speculate on its resolution and interpretation.

We further note that the same behavior occurs in the case of the dynamical scalar model considered in  Section~\ref{sec:gwdriven}, although the kink feature is absent.

\subsection{Instabilities}

Here we briefly examine the stability of fluctuations for different values of the position of the IR brane.  Summarizing the results first:  the minimum for smallest $z_\textnormal{c}$ and neighboring values of the radion VEV is always a ``healthy" minimum where there are no unresolved tachyonic states.  However, past the first region where the Higgs VEV resolves the tachyon, there are apparently instabilities without condensates to rectify them, or the condensates are insufficient to prevent all of them.  In this case, the approximation of a static 5D description is not a good one, and the theory must resolve the tachyon with some more dramatic dynamics.  We comment briefly on this  in Section~\ref{sec:discussion}, and a more complete analysis of this region is part of future work.

We can inspect the tachyon instabilities through examination of the spectrum of Higgs fluctuations.  The equation of motion for these, presuming a vanishing Higgs VEV, is given by
\beq
h''(z) -\frac{3}{z} h'(z) -\frac{1}{z^2} \left(-4+\delta m^2-\lambda z^\epsilon\right) h(z)= - m^2 h(z),
\label{eq:fluceom}
\eeq
with IR boundary condition given by
\beq
h'(z_1) = \frac{1}{2 z_1} \lambda_H v_H^2 h(z_1).
\eeq
The UV boundary condition we impose is that the solution must asymptote to the behavior given in Eq.~(\ref{eq:hvevsol}), on the $+$ branch.  Equivalently, one could impose $z_0h'(z_0) = m_0^2 h(z_0)/2$ for some non-tuned value of $m_0^2$, and subsequently take the limit as $z_0 \rightarrow 0$.  The result is similar.

\begin{figure}[t!]
	\centering
	\includegraphics[width=.7 \textwidth]{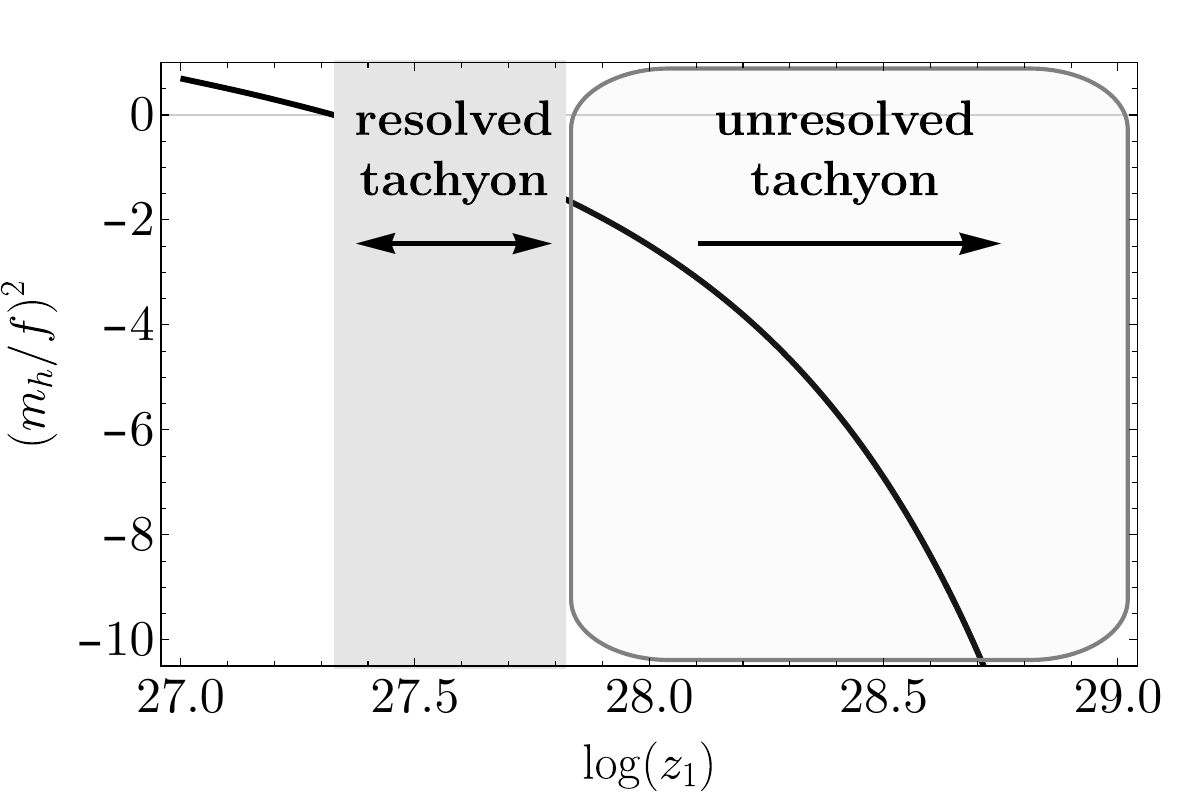}
	\caption{\small Here we show the lowest eigenvalue associated with the Higgs fluctuations, the solutions to Eq.~(\ref{eq:fluceom}) with the boundary conditions associated with the IR brane-localized Higgs potential.  The region where the Higgs VEV resolves a single tachyon is shaded, and the physical Higgs fluctuation here is in fact massive.  This is the first critical region, where there is only one tachyon to be resolved.  An unresolved tachyon emerges for larger $z_1$, when the Higgs VEV turns off, indicating a fundamental instability.}
	\label{fig:higgstach}
\end{figure}

Solving the boundary value problem gives the spectrum of states.  In Figure~\ref{fig:higgstach}, we display the lowest eigenvalue, expressed as the ratio $(m_h/f)^2$, where $f^{-1} = z_1$.  The region where the Higgs VEV resolves the tachyon is shaded, and there Higgs fluctuations are massive.  At larger $z_1$, outside the gray region where there is no Higgs condensate, there is a tachyon that persists.   It appears that the condensate can rectify the tachyon so long as the mode is not \emph{too} tachyonic.  For larger $z_1$, the problem grows still worse:  in the neighborhood of the $n$-th critical value of $z_1$, there are $n-1$ unresolved tachyons.

Due to the presence of unresolvable tachyons for larger $z_1$, we focus our main attention on values of $z_1$ where there is either no tachyon, at $z_1 < z_\textnormal{c}^1$, or there is a single tachyon that is resolved by the vacuum expectation value for the Higgs field, $z_1 \gtrsim z_\textnormal{c}^1$.

It is interesting that the first minimum appears metastable, with other minima at larger values of $z_1$, but lower effective potential energy, as seen in Figure~\ref{fig:dilpot}.   We comment further on this in Section~\ref{sec:discussion}.  However, we must not take this region of large $z_1$ too seriously.  For one, the vacuum expectation value for the Higgs diverges at the end of the first condensate region, and the theory cannot be trusted there, where gravitational backreaction will be very large.  It is not obvious that the geometry can be considered at all past this point.

Additionally, as discussed in the previous subsection, we have not taken into account the required time dependence in the solution.  It is not obvious that the same instabilities will be present once the ansatz for the background is relaxed to allow for a dynamical background.  

Finally, we have enforced a rigid dependence of the mass on the $z$-coordinate, which is not likely to be possible in a self-contained fully dynamical model.  It is likely that this rigidity of the mass profile is responsible for the kinked behavior of the potential at the critical point.  Backreaction between the dynamics that drives the mass and the Higgs VEV may smooth out the potential.  This is confirmed in a fully dynamical model explored in the next section.\footnote{We thank Prashant Saraswat and Michael Geller for useful comments on the first version of this pre-print that led us to more fully explore the details of the dynamical model.}

\section{Dynamical Model}
\setcounter{equation}{0}
\label{sec:gwdriven}

In the previous section, the varying bulk mass for the Higgs was taken as an input, breaking the isometries of AdS explicitly, and we took the UV brane to the AdS boundary.  In this section we show how similar physics can be derived in a dynamical and more realistic model with a UV brane where the interplay of a Goldberger-Wise-like scalar field and a bulk Higgs on an AdS background generates a similar physics result.  This model has the benefit of being fully dynamical, of having a possible CFT dual that is easier to interpret, and having a modulus potential that does not exhibit the kink of the previous toy model.  The 5D Higgs profile can no longer be solved for analytically, and so we perform a numerical study.  

This model has two bulk scalar fields, a real scalar, $\phi_\textnormal{d}$, the ``driving scalar", in addition to a Higgs scalar.   The fields are coupled in such a way that a varying VEV of the scalar $\phi_\textnormal{d}$ drives the effective bulk mass of the Higgs, making it a function of the extra-dimensional coordinate.  The 5D action is given by
\beq
\begin{aligned}
	S &= \int d^4x dz \sqrt{g} \left[  |\partial_M H|^2 + \frac{1}{2} (\partial_M \phi_\textnormal{d})^2 +\frac{6}{\kappa^2}-\left(m^2_H-\lambda \phi_\textnormal{d} \right)|H|^2 -\frac{1}{2} m_{\phi_\textnormal{d}}^2 \phi_\textnormal{d}^2 -\frac{1}{2 \kappa^2}R \right] \\
	&\phantom{{}={}}- \int d^4 x  z^{-4} V_0 (\phi_\textnormal{d}, |H|)\bigg\rvert_{z\rightarrow z_0} -\int d^4 x z^{-4} V_1 (\phi_\textnormal{d},|H|)\bigg\rvert_{z\rightarrow z_1}.
\end{aligned}
\eeq
The brane potentials are assumed to take the form $V_{0,1} = \delta T_{0,1} + V_{0,1}^{\phi_\textnormal{d}} + V_{0,1}^{H}$.
For the brane Higgs potentials, we take 
\beq
V^H_0 = m_0^2 |H|^2, \qquad \qquad V^H_1 = \lambda_H |H|^2 \left(|H|^2-v_H^2\right).
\eeq
The potential here is quite similar to that in~\cite{Graham:2015cka}, however in this model, there are no very small parameters, and this is a 5D bulk potential -- the effective bulk potential in the 4D theory is of course related, but not in one-to-one correspondence.  There is a mild approximate shift symmetry in the bulk for the $\phi_\textnormal{d}$ scalar -- a slightly suppressed bulk mass parameter as utilized in the Goldberger-Wise stabilization mechanism~\cite{Goldberger:1999uk}, and a perturbative bulk interaction with the 5D $H$ field.

The equation of motion for $\phi_\textnormal{d}$ is given by
\beq
\phi''_\textnormal{d} - \frac{3}{z} \phi'_\textnormal{d} - \frac{1}{z^2} \left( m_{\phi_\textnormal{d}}^2 \phi_\textnormal{d} - \frac{\lambda}{2} \phi_h^2 \right) =0,
\label{eq:phideom}
\eeq
which is a simple linear non-homogeneous equation for a given Higgs background, $\phi_h$, and the solution is given by
\beq
\phi_\textnormal{d} =  z^\epsilon \left( \phi_\epsilon + \frac{\lambda}{4 (2 - \epsilon)} \int_{z_0}^{z} \phi_h^2(\tilde{z}) \tilde{z}^{-1-\epsilon} d\tilde{z} \right) +  z^{4-\epsilon} \left( \phi_4 - \frac{\lambda}{4 (2-\epsilon)} \int_{z_0}^{z} \phi_h^2(\tilde{z}) \tilde{z}^{-5+\epsilon} d\tilde{z} \right),
\label{eq:phidsol}
\eeq
where $\epsilon = 2 - \sqrt{4 + m_{\phi_\textnormal{d}}^2}$.  We take $\epsilon$ to be small and positive, but not tiny, e.g.\ $\epsilon = {\mathcal O}(0.1)$, corresponding to a small tachyonic bulk mass for $\phi_\textnormal{d}$.  There are two contributions to the solution -- the solution to the homogeneous part of Eq.~(\ref{eq:phideom}), and the integrals that solve the nonhomogeneous part from a non-vanishing Higgs VEV.

The Higgs equation of motion is given by
\beq
\phi''_h - \frac{3}{z} \phi'_h - \frac{1}{z^2} \left( m_H^2 - \lambda \phi_\textnormal{d} \right) \phi_h = 0,
\eeq
which can, in principle, be expressed as a single nonlinear integro-differential equation by inserting the  solution to the $\phi_\textnormal{d}$ boundary value problem.

The radion potential, again assuming small gravitational backreaction, now takes contributions from both fields, and is given in general by
\beq
\begin{aligned}
	V_\text{rad} &= \frac{1}{z_0^4} \left[ V_0 - \frac{6}{\kappa^2} \sqrt{G_0}\right] + \frac{1}{z_1^4} \left[ V_1 + \frac{6}{\kappa^2} \sqrt{G_1}\right] \\
	&\approx  \frac{1}{z_0^4} \left[ \delta T_0 + V_0^{\phi_{\textnormal{d},0}} + V_0^{H} + \frac{1}{4} m_H^2 (z_0) \phi_{h,0}^2 - \frac{1}{4} z_0^2 {\phi'}_{h,0}^2+ \frac{1}{4} m_{\phi_{\textnormal{d}}}^2 \phi_{\textnormal{d},0}^2 - \frac{1}{4} z_0^2 {\phi'}_{\textnormal{d},0}^2 \right]  \\
	&\phantom{{}={}} + \frac{1}{z_1^4} \left[ \delta T_1 + V_1^{\phi_{\textnormal{d},1}} + V_1^{H} - \frac{1}{4} m_H^2 (z_1) \phi_{h,1}^2 + \frac{1}{4} z_1^2 {\phi'}_{h,1}^2- \frac{1}{4} m_{\phi_\textnormal{d}}^2 \phi_{\textnormal{d},1}^2 + \frac{1}{4} z_1^2 {\phi'}_{\textnormal{d},1}^2 \right].
\end{aligned}
\label{eq:radpotGW}
\eeq
Using the analytic solution for $\phi_\textnormal{d}$ in terms of the function $\phi_h$ from Eq.~(\ref{eq:phidsol}), supplemented by the boundary conditions for $\phi_\textnormal{d}$, one can express the entire radion potential purely in terms of the solution to the $\phi_h$ equation of motion.

We first specify to a model where much can be done analytically to show some basic results in this model.  We take the IR brane potential for the $\phi_\textnormal{d}$ scalar to be a localized mass term, $V_{1}^{\phi_\textnormal{d}} = - \epsilon \phi_\textnormal{d}^2$.  If we assume this value for the mass, and if the Higgs VEV is taken to be zero, the solution is $\phi_\textnormal{d} = v_0 (z/z_0)^{\epsilon}$.  Without the Higgs contribution, the radion potential is just a scale-invariant quartic.  We then assume a UV brane potential that fixes $v_0$ to some value (a stiff-wall-type boundary condition).   In this background, ignoring the nonhomogeneous part of $\phi_\textnormal{d}$ in the Higgs equation of motion, the solution for the Higgs VEV is the same as described in the previous section.

Also, in the case of the model under consideration, there are considerable simplifications of the radion potential.  Specifically, all of the $\phi_\textnormal{d}$ terms in the IR brane contribution cancel after imposing the boundary condition $z \phi_\textnormal{d}' = \epsilon \phi_\textnormal{d}$, which arises from the boundary potential we have chosen for $\phi_\textnormal{d}$.

The radion potential, expressed purely in terms of the boundary values of and integrals over the solution to the Higgs equation of motion, is
\beq
\begin{aligned}
	V_\textnormal{rad} &=  \frac{1}{z_0^4} \left[ \delta \tilde{T}_0 + \frac{m_0^2}{2} \phi_{h,0}^2 + \frac{1}{4} \left( m_H^2 - \lambda v_0 \right) \phi_{h,0}^2 - \frac{1}{4} \left( z_0 \phi'_{h,0} \right)^2 - \frac{1}{2} \epsilon v_0 z_0^{4-\epsilon} I_4 - \frac{1}{4} \left( z_0^{4-\epsilon} I_4 \right)^2 \right]  \\
	&\phantom{{}={}} +   \frac{1}{z_1^4} \left[ \delta \tilde{T}_1  - \frac{1}{4} \left\{ m_H^2 - \lambda \left( \frac{z_1}{z_0} \right)^\epsilon \left( v_0 + \frac{z_0^\epsilon}{2 (2-\epsilon)} I_\epsilon - \frac{z_0^{4-\epsilon}}{2 (2-\epsilon)} I_4 \right) \right\} \phi_{h,1}^2 \right. \\
	&\phantom{{}={}} +  \left.\frac{1}{4} \lambda_H \phi_{h,1}^2 \left( \phi_{h,1}^2-2 v_H^2 \right) + \frac{1}{4} \left( z \phi'_{h,1} \right)^2 \right],
	\label{eq:effpotsimpdriving}
\end{aligned}
\eeq
where $I_\epsilon$ and $I_4$ are the following integrals:
\beq
\begin{aligned}
	I_\epsilon = &\frac{\lambda}{2} \int_{z_0}^{z_1} \phi_h^2 (\tilde{z}) \tilde{z}^{-1-\epsilon} d\tilde{z}, \\
	I_4 = & \frac{\lambda}{2} \int_{z_0}^{z_1} \phi_h^2 (\tilde{z}) \tilde{z}^{-5+\epsilon} d\tilde{z}.
\end{aligned}
\eeq
This expression is exact up to contributions from gravitational backreaction, which we are here neglecting.  

We note that one can show analytically by expanding the solution near the critical point that the kink of the previous section is removed by the backreaction of the Higgs VEV onto the driving scalar.  We provide the calculation in Appendix~\ref{app:kink}.

In the limit of tiny Higgs VEV, the Higgs background is well approximated by the solution that assumes $\phi_\textnormal{d} = v_0 (z/z_0)^\epsilon$, which was explored in the previous section:
\beq
\phi_h = z^2 \left( \phi_+ J_\nu \left(z^{\epsilon/2} \frac{2\sqrt{\lambda v_0}}{\epsilon} \right) + \phi_- J_{-\nu}  \left(z^{\epsilon/2} \frac{2\sqrt{\lambda v_0}}{\epsilon} \right) \right),
\eeq
where $\nu = 2\sqrt{\delta m^2}/\epsilon$, with $\delta m^2 = m_H^2 +4$.  This is sufficient for determining the critical point, however, for the purposes of evaluating the effective potential, a full numerical solution is necessary.  That is, past the critical point, the Higgs background looks similar to this profile, but the differences due to the nonlinearities need to be incorporated in order to correctly determine the radion potential.

In the next subsection, we embark on a full numerical analysis of the coupled equations, working with a more common setup for the $\phi_\textnormal{d}$ scalar, where its boundary values are set to $v_{0,1}$ on the UV/IR branes.

\subsection{Stiff Wall Model}
In the stiff wall limit, the boundary conditions for $\phi_\textnormal{d}$ are $\phi_\textnormal{d}(z_{0,1}) = v_{0,1}$.  Thus, the value of the bulk Higgs mass varies from $m_H^2(z_0)  = m_H^2 - \lambda v_0$ on the UV brane to $m_H^2(z_1)  = m_H^2 - \lambda v_1$ in the IR.  It is not difficult to arrange for the effective Higgs mass to vary such that it crosses the BF bound somewhere in the bulk, and evolves from power law behavior in the UV to a log-periodic power law in the IR.  For small values of $z_1$, the AdS tachyon is not reached, however, for larger values, the tachyon eventually must emerge and be resolved by condensation of some sort.  In order to understand this process fully, we embark on a numerical exploration of solutions to the scalar equations of motion.

We note that the bulk equations governing the behavior of $\phi_\textnormal{d}$ and $\phi_h$ are nonlinear, and that one must take care in seeking out solutions to the scalar equations of motion.  Existence and uniqueness are not  guaranteed in the case of the general nonlinear boundary value problem.   Due to this complication, we do not search for solutions with fixed $z_1$, but rather search for solutions with varying values of the Higgs VEV.  We note that there is always a solution with $\phi_h = 0$ -- the solution to the linear Goldberger-Wise problem with just the $\phi_\textnormal{d}$ scalar.  

To numerically investigate the solutions, we generalize the bulk equations to accommodate a sort of shooting method.  That is, we search for solutions to the boundary value problem at hand by scanning over a range of initial value problems until we find a global solution.   To do so, we write the solution to the Higgs as $\phi_h = h_0 f_h (z)$ with $f_h(z_0)$ set to some arbitrary value.    The coefficient $h_0$ does not appear explicitly in the Higgs equation of motion, as for fixed $\phi_\textnormal{d}$, that equation is linear.  It does, however, appear in the $\phi_\textnormal{d}$ equation:
\beq
\begin{aligned}
	& f_h'' -\frac{3}{z} f_h' -(m_H^2 - \lambda \phi_\textnormal{d} ) f_h =0, \\
	& \phi_\textnormal{d}''-\frac{3}{z} \phi_\textnormal{d}' - (\epsilon^2 - 4 \epsilon) \phi_\textnormal{d} + \frac{\lambda}{2} h_0^2 f_h(z)^2 = 0.
\end{aligned}
\eeq
We shoot from the UV brane, enforcing the UV brane boundary conditions for $f_h$ and $\phi_\textnormal{d}$, which in our cases of study are $f'_h (z_0) = m_0^2 f_h (z_0)/(2 z_0)$ and $\phi_\textnormal{d}(z_0) = v_0$.  We then use a prechosen value of $h_0$ to define each solution.  The magnitude of $h_0$ closely tracks the low energy effective Higgs VEV as a fraction of the KK scale.   As $f_h (z_0)$ is fixed arbitrarily, the only remaining condition is on $\phi_\textnormal{d}'$, the correct value for which will be determined by shooting.  The value of $z_1$ for a $\phi_\textnormal{d}'$ guess is chosen by finding the point in the bulk at which the IR brane boundary condition for the Higgs is met:  $h_0^2 = 1/f_h(z)^2 \left( v_H^2 - \frac{2 f'_h (z) }{\lambda_H z f_h(z)} \right)$.  Of course at that $z_1$, it will not usually be the case that the IR boundary condition for $\phi_\textnormal{d}$ is met, so we repeat the process by shooting with different values of $\phi_\textnormal{d}'$ until both the Higgs and $\phi_\textnormal{d}$ boundary conditions are solved at the same value of $z_1$.  In summary, for a given input Higgs VEV, we obtain a value for $z_1$ and the associated bulk profiles for $\phi_h$ and $\phi_\textnormal{d}$.

It is useful to characterize each solution in terms of some parameter with physical meaning to a low energy observer, and we choose the effective Higgs VEV, evaluated by integrating the Higgs solution squared over a hypothetical flat zero-mode gauge boson wave function with appropriate metric factors:
\beq
\left( \frac{v_\text{eff}}{f} \right)^2 = z_1^2 h_0^2 \int_{z_0}^{z_1} \frac{1}{z^3} f_h^2(z) dz.
\eeq

We find that the types of solutions one obtains can be divided into two classes based on the value of the IR brane-localized Higgs mass, determined by taking different values for $v_H^2$.  If $v_H^2$ is taken to be very negative, corresponding to a non-tachyonic brane-localized Higgs mass squared, there is no solution for nonzero Higgs VEV with $z_1 > z_\textnormal{c}$.  Instead, with increasing Higgs VEV (roughly $h_0$ in our numerical analysis), the solution requires \emph{smaller} values of $z_1$.  This then means that for a given value of $z_1 < z_\textnormal{c}$, there may be two solutions, one with positive effective 4D Higgs mass squared/vanishing VEV, and the other with nonvanishing Higgs VEV.  For more positive values of $v_H^2$, a different behavior is possible, e.g. where the value of $z_1$ at first increases as the Higgs VEV is turned on, and then turns around, so that at some particular $z_1$ value, there are two possible values of the effective Higgs VEV.  

\begin{figure}[t]
	\centering
	\includegraphics[height=2in]{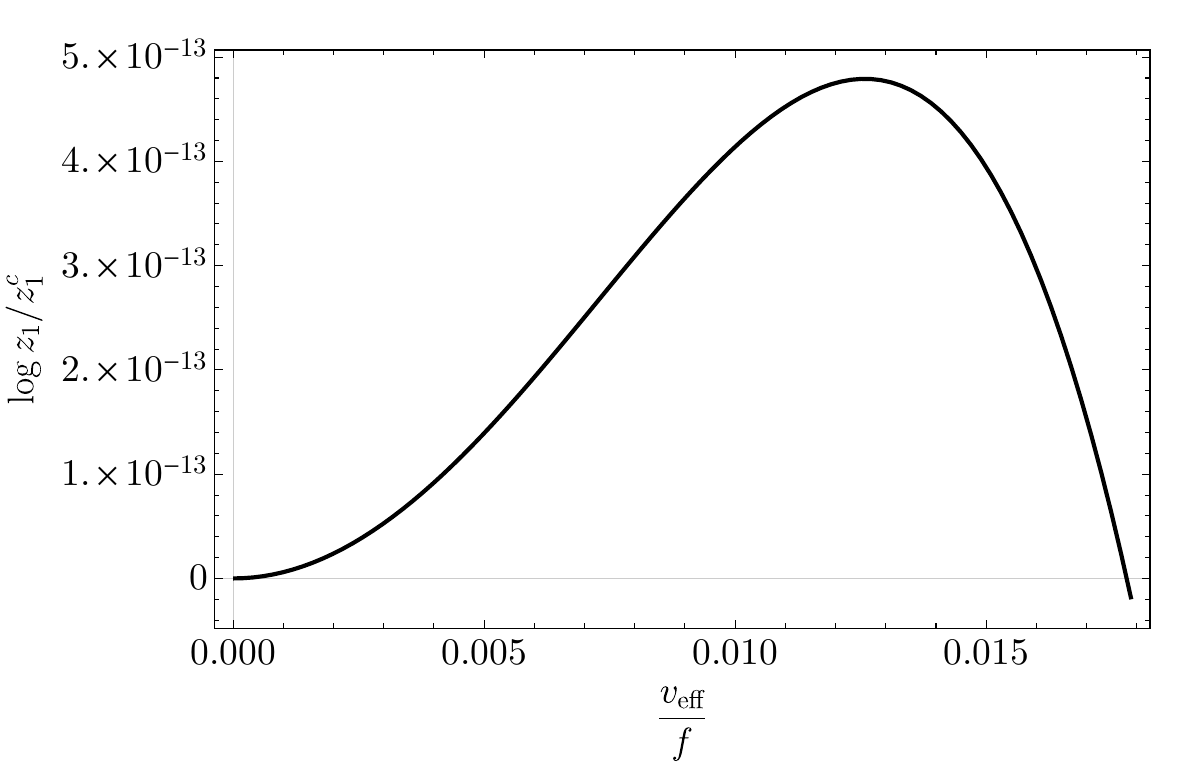}\includegraphics[height=2in]{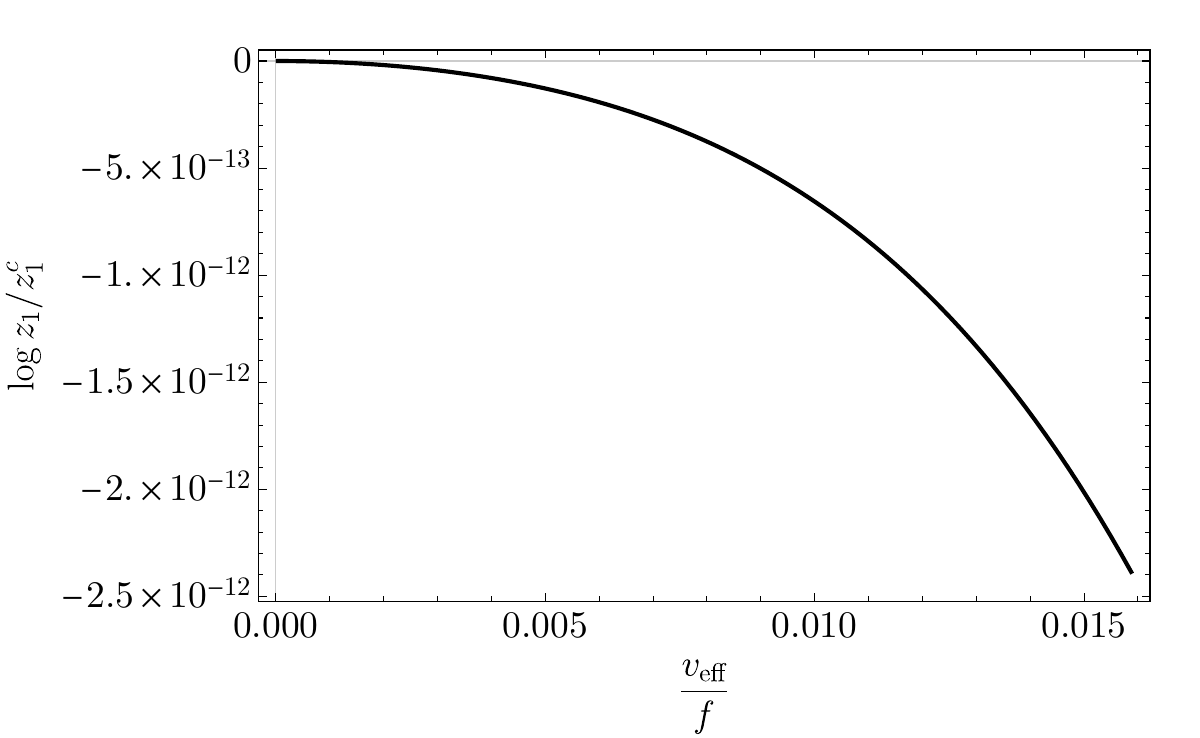}
	\includegraphics[height=1.75in]{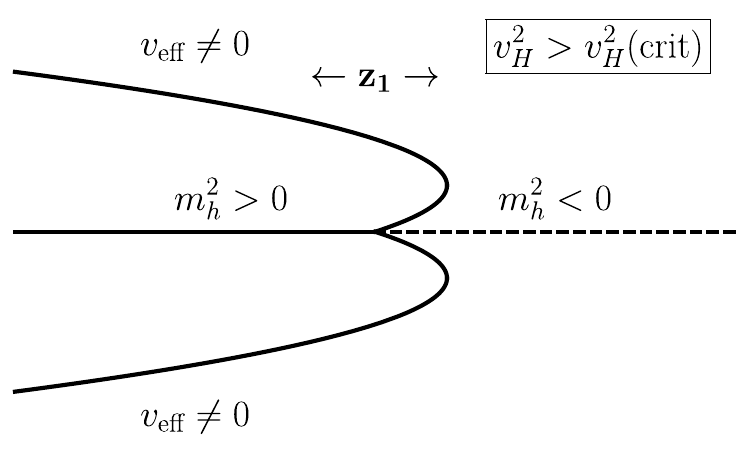}\includegraphics[height=1.75in]{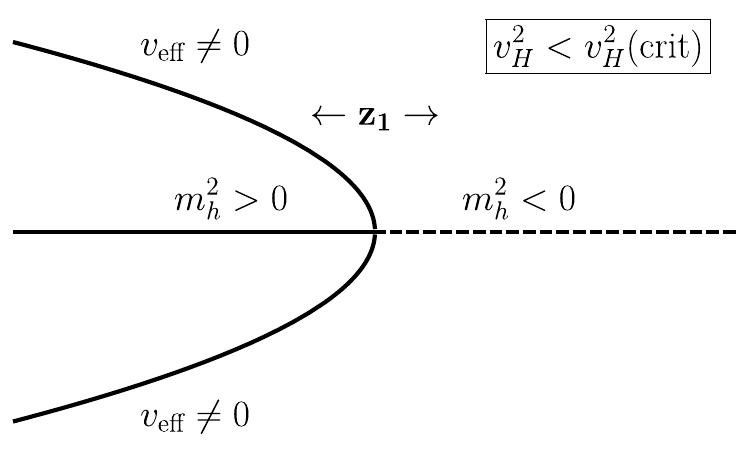}
	\caption{\small In this plot, we show the effective VEV as a function of $z_1$.  On the left, we show it for a  value of $v_H^2$ that is very close to ``critical," but with $v_H^2 > v_H^2 (\text{crit})$.  In this case, for small Higgs vev, $z_1 > z_c$.  On the right, we display it for $v_H^2$ more negative than the critical value, and in this case, for all values of the Higgs VEV, we find $z_1 < z_c$.  We also sketch the ``bifurcation" diagrams for each of these scenarios as a function of $z_1$, where the solid lines represent the stable scalar configurations and the dashed line represents the background solution with unresolved tachyon(s).  The branching point corresponds to $z_1 = z_\textnormal{c}$.}
	\label{fig:veffplot}
\end{figure}

None of this should be too surprising from the perspective of the 4D CFT dual.  The value of $z_1$ represents some total scale of conformal breaking through the relation $z_1 \sim 1/f$ where $f$ should be some additive combination of all vacuum expectation values in the system that break approximate conformal invariance spontaneously.  The bulk $\phi_\textnormal{d} |H|^2$ interaction has enforced a relationship between the multiple $f$'s (in this case, dual to the $\phi_\textnormal{d}$ and $\phi_h$ scalar solutions), and lifted the constraint of uniqueness present in theories without bulk scalar interactions.  In other words, the nonlinearity of the boundary value problem has allowed for multiple solutions for a given $z_1$, and thus different ways for the bulk theory to produce a given total $f \sim 1/z_1$.  In Figure~\ref{fig:veffplot}, we show for two different values of $v_H^2$ that are close to a critical value (but on either side) the relationship between $z_1$ and the effective VEV.  We note that the Higgs VEV grows extremely fast near the critical point.

This matches onto the first part of our preliminary discussion of the dilaton effective theory in Section~\ref{sec:eft}.   Extremization of the scalar part of the action corresponds in the dual picture to sorting out the correct relationship between the vacuum expectation values of operators in the approximate CFT.  This relationship must now be fed into the total radion effective potential.

\begin{figure}[t!]
	\centering
	\includegraphics[width=0.5\textwidth]{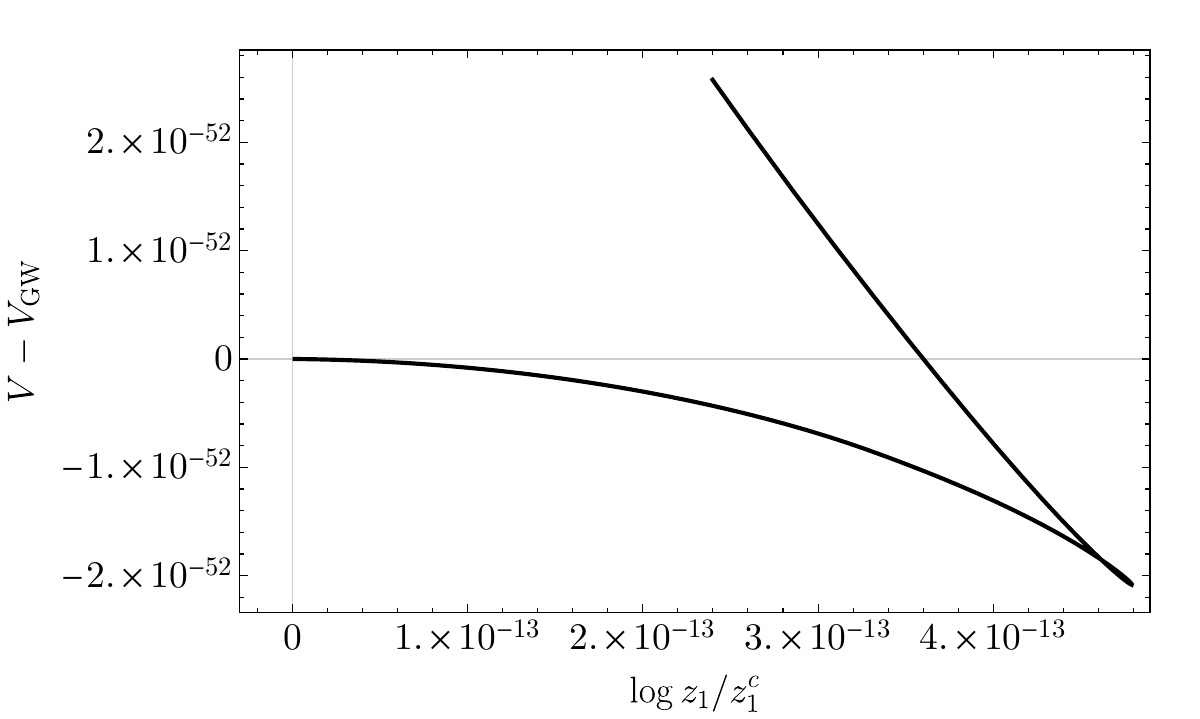}\includegraphics[width=0.5\textwidth]{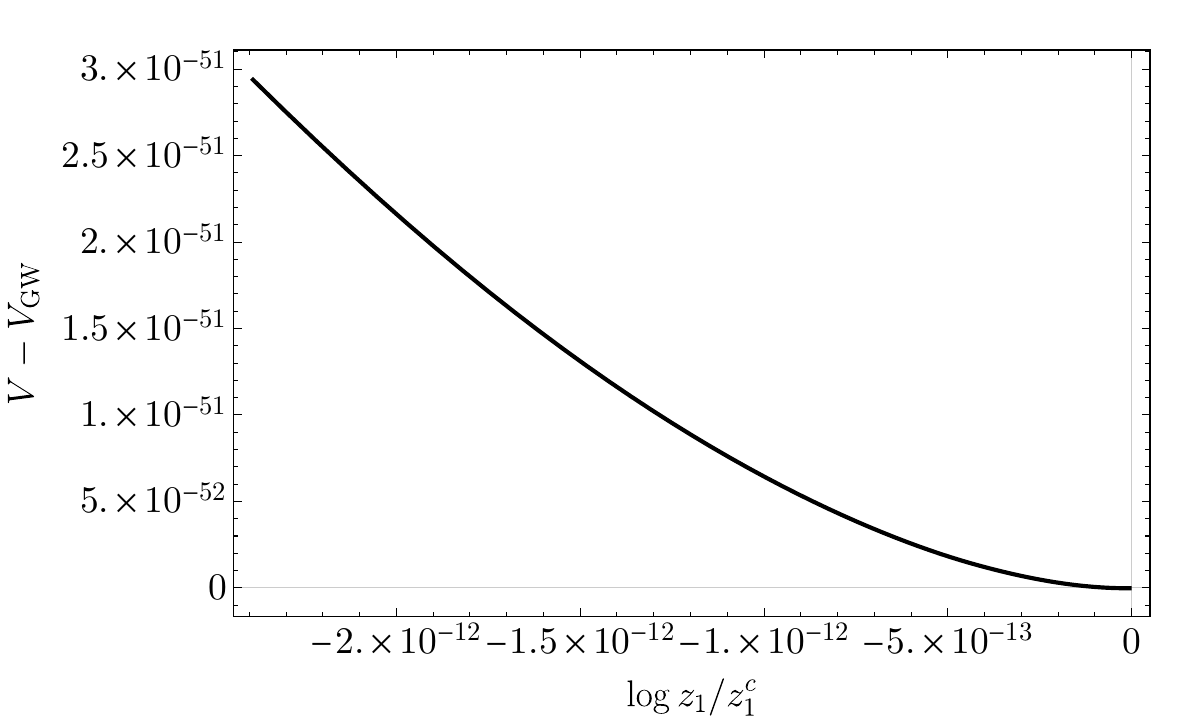}
	\includegraphics[width=0.5\textwidth]{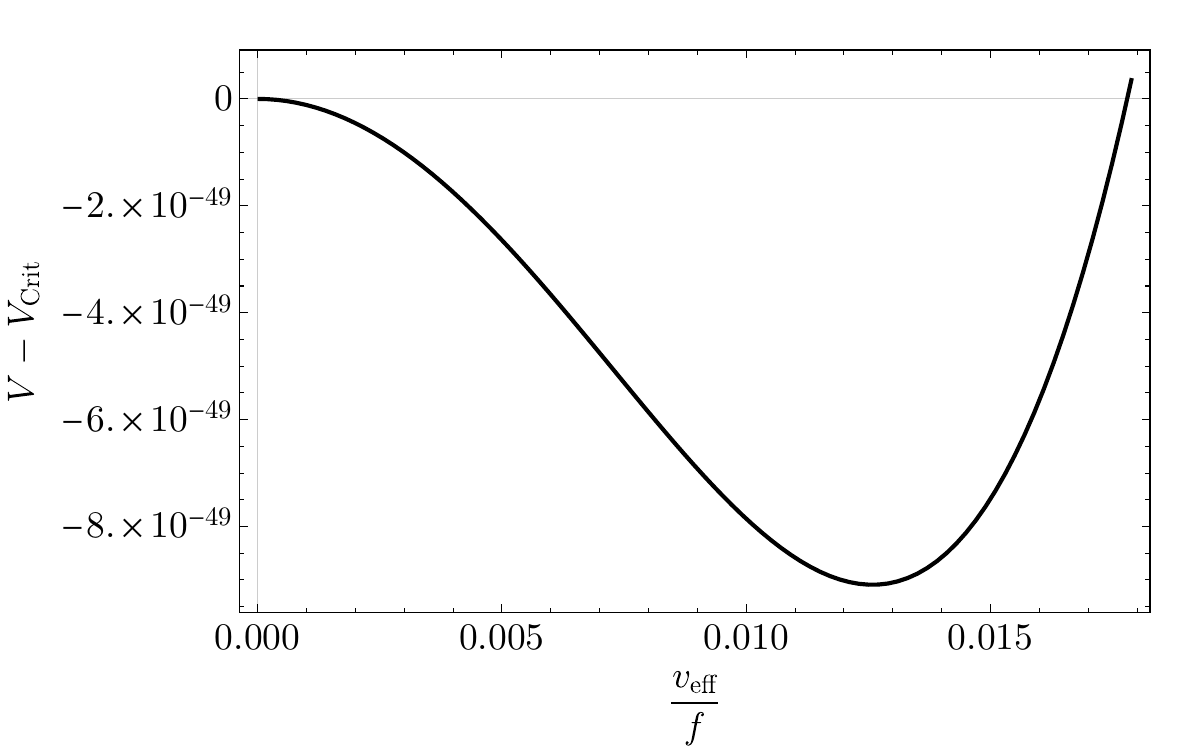}\includegraphics[width=0.5\textwidth]{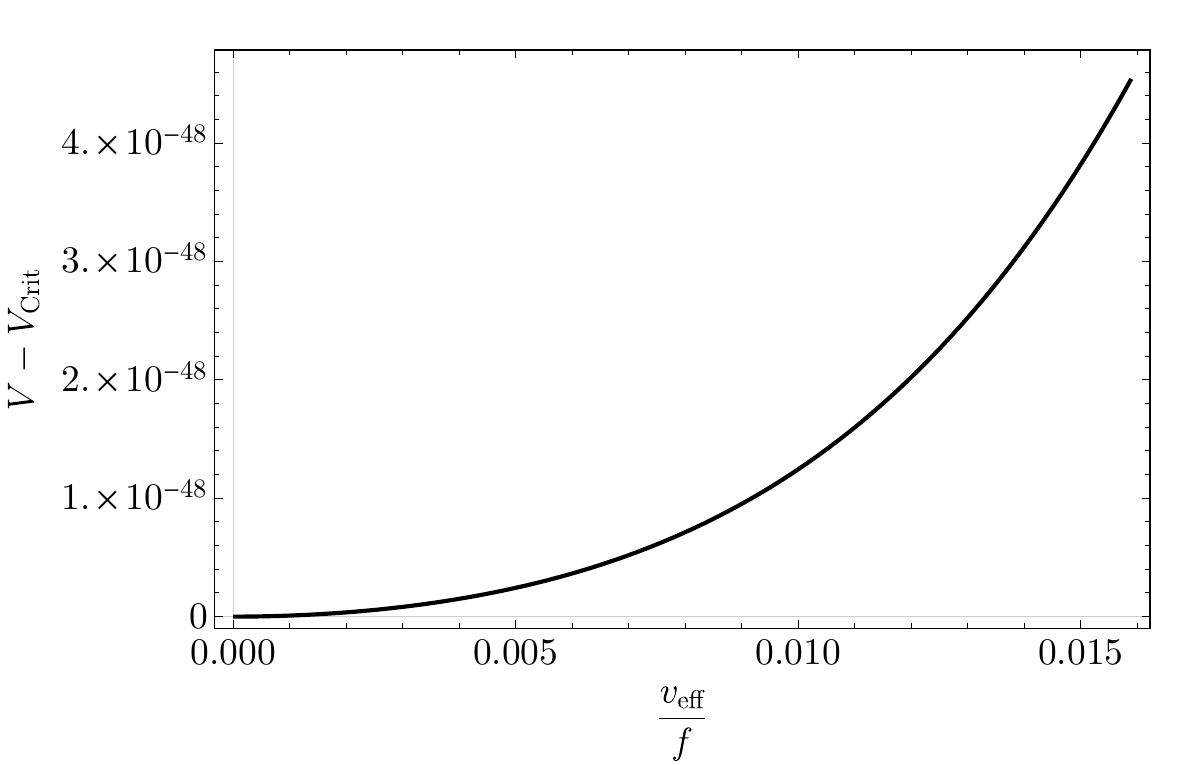}
	\includegraphics[width=0.5\textwidth]{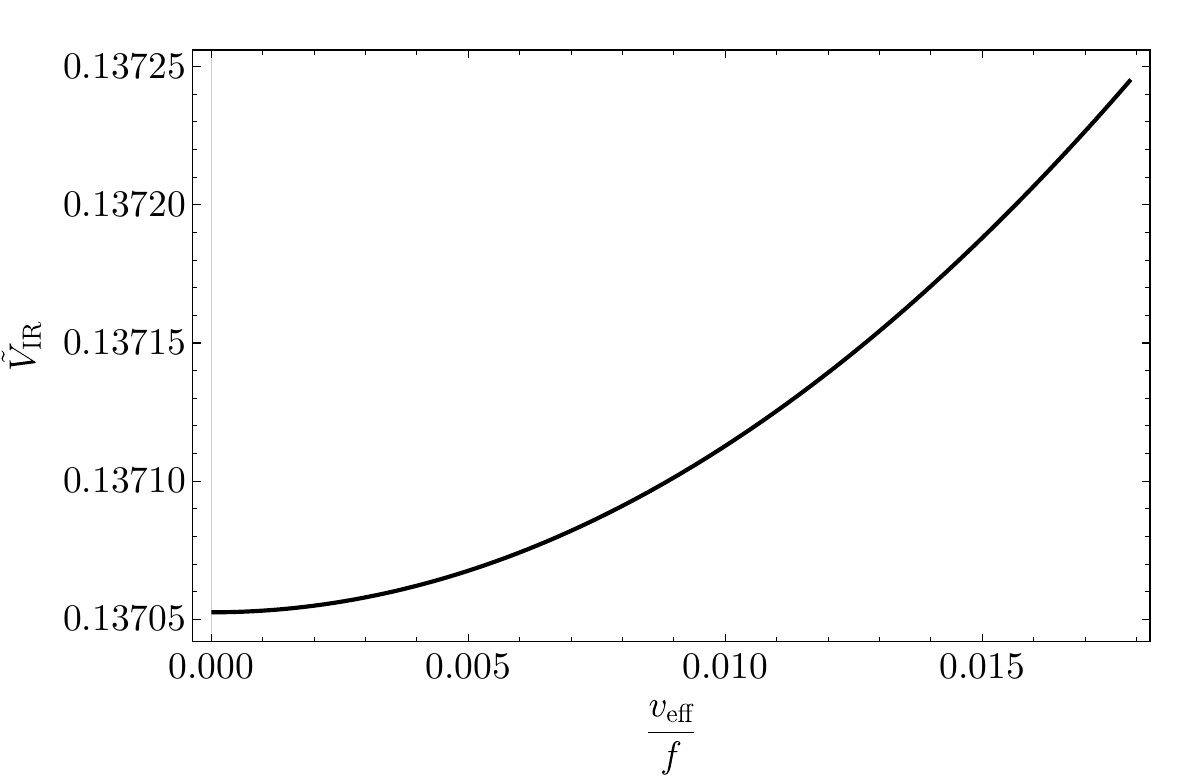}\includegraphics[width=0.5\textwidth]{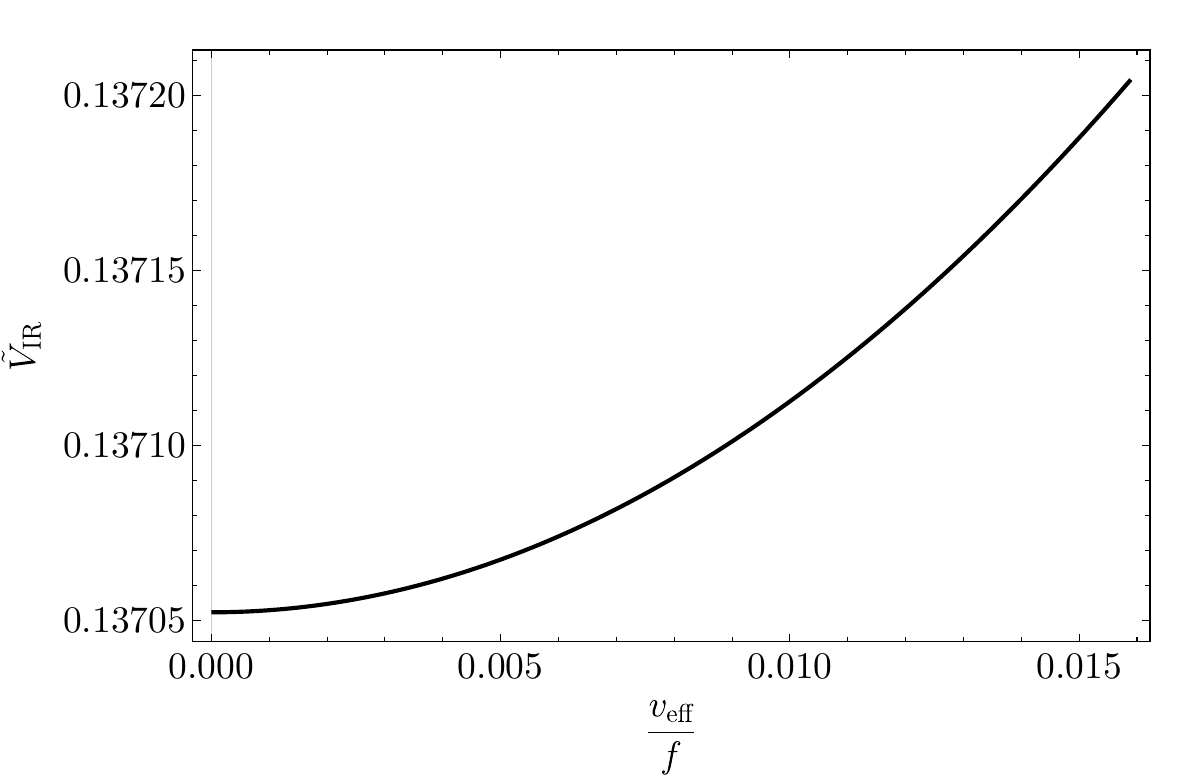}
	\caption{\small In this plot, we show the radion potential for two different values of $v_H^2$ near $v_H^2 (\text{crit})$.  In the column on the left, $v_H^2 > v_H^2 (\text{crit})$.  We have taken $\epsilon = v_0 = 1/10$, $m_H^2 = -3.9$, $v_0 = 1$, $\delta T_1 = -1/10$, $\lambda = 1/3$, $m_0^2 = 4.1$, and $\lambda_H = 1/8$.  The critical point is between $v_H^2 = -16.5830$ and $v_H^2 = -16.5831$, and the two columns correspond respectively to these two values of $v_H^2$ that straddle $v_H^2 (\text{crit})$.  In descending order, the plots display:  the difference between the radion potentials with and without a Higgs VEV as a function of $\log z_1/z_1(\text{crit})$, the same potential with $V_{\text{crit}}$ being its value at the critical $z_1$, but instead as a function of $v_\text{eff}/f$, and finally the value of $\tilde{V}_\text{IR}$, defined in Eq.~(\ref{eq:vtildes}), indicating the degree of mismatch of the metric junction condition on the IR brane.  There is no discernible difference in these two plots on either side of the critical value of $v_H^2$, and there is certainly no zero.}
	\label{fig:radpotfull}
\end{figure}

Now we consider the radion potential for this model.  Plugging in the results from the numerical solutions, we find that for $v_H^2 > v_H^2 (\text{crit})$, there are solutions to the scalar equations of motion and boundary conditions with $z_1 > z_\textnormal{c}$.  For all $v_H^2 < v_H^2 (\text{crit})$, there is a minimum in the effective radion potential \emph{at $z_1=z_\textnormal{c}$}, the value of $z_1$ where the Higgs fluctuation is massless.  We note that this is under the constraint that values of $z_1$ where there is an unresolved tachyon are disallowed.
\footnote{Dynamics of the system \emph{will} move the IR brane into this region - the true vacuum is not simply one in which the brane rests at $z_c$.  Going past the critical value of $v_H^2$, there is likely a phase transition of the system that cannot be encapsulated under the static ansatz that was the starting point for this analysis.  We speculate on its nature in Section~\ref{sec:cosmo}, leaving a full analysis of the $v_H^2 < v_H^2 (\text{crit})$ region for future work.}

    In Figure~\ref{fig:radpotfull}, we display the results of the radion potential for two values of $v_H^2$ near the critical value.  We show the potential both as a function of $z_1$ and as a function of $v_\text{eff}/f$. In this plot, $V_{\text{GW}}$ denotes the GW contribution to the radion potential, while $V_{\text{crit}}$ is the radion potential at the critical $z_1$.  For the subcritical case, the location of the potential minimum is at an energy which is lower than if the Higgs VEV is turned off -- symmetry breaking is the preferred configuration for that value of $z_1$.  

\begin{figure}[t]
	\centering
	\includegraphics[width=0.66\textwidth]{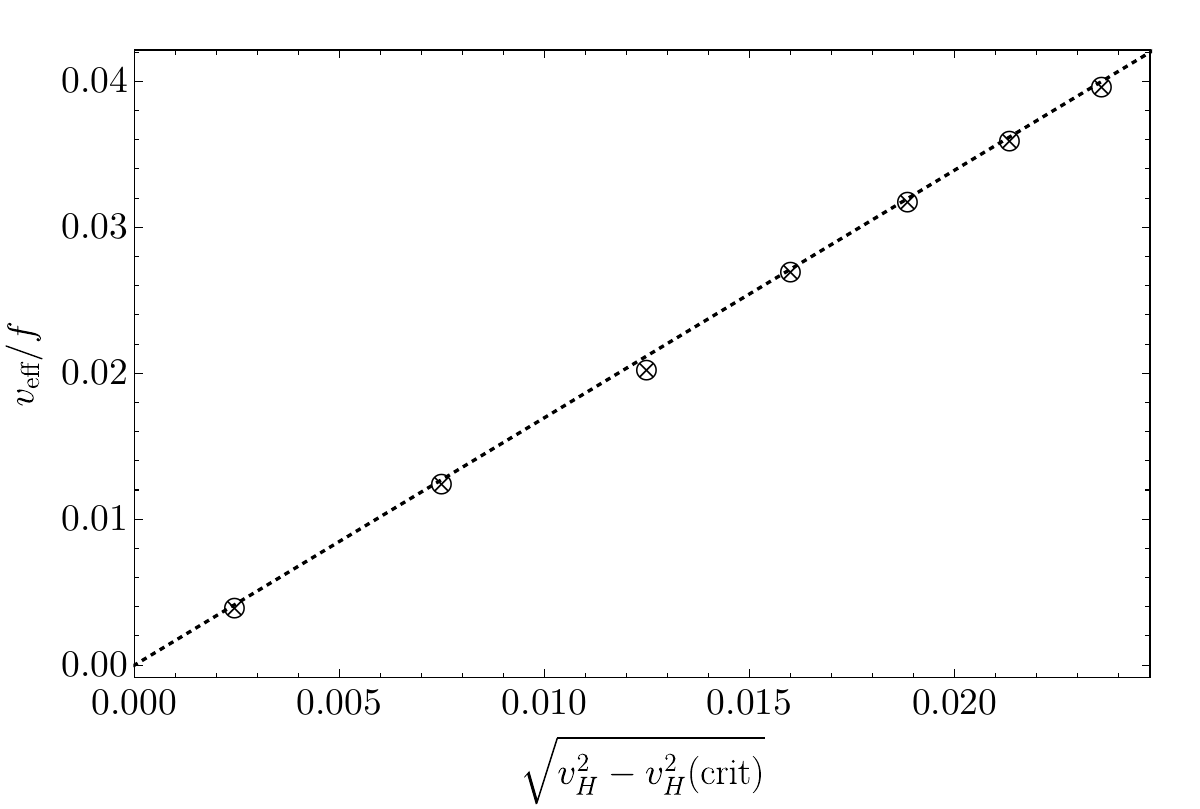}
	\caption{\small In this plot, we display the behavior of $v_\text{eff}/f$ at the minimum of the radion potential for subcritical $v_H^2 > v_H^2 (\text{crit})$ as it approaches the critical region.  The dashed line is a linear fit to the numerical data forced to pass through the origin by adjusting $v_H^2(\text{crit})$.  
	The critical value is determined in this manner to be $v_H^2(\text{crit}) = -16.58305605$
	}
	\label{fig:vhcrit}
\end{figure}

One can now make contact with the dilaton effective potential discussed in Section~\ref{sec:eft}.  The value of $1/z_1$ for a given solution corresponds to the total scale of symmetry breaking, with the KK-mode masses of the extra-dimensional theory corresponding to its value.    Larger values of $z_1$ have scalar instabilities.  Ignoring this region, the radion potential is minimized at the largest value of $z_1$ that accommodates a solution to the scalar equations of motion.

In Figure~\ref{fig:vhcrit}, we display the behavior of the model with subcritical $v_H^2$, focusing on the value of the Higgs VEV in the approach to Higgs criticality.  The Higgs VEV, roughly the inverse correlation length in the low energy theory, appears to depend linearly on $\sqrt{v_H^2-v_H^2(\text{crit})}$ in the approach.

\begin{figure}[t]
	\centering
	\includegraphics[width=0.5\textwidth]{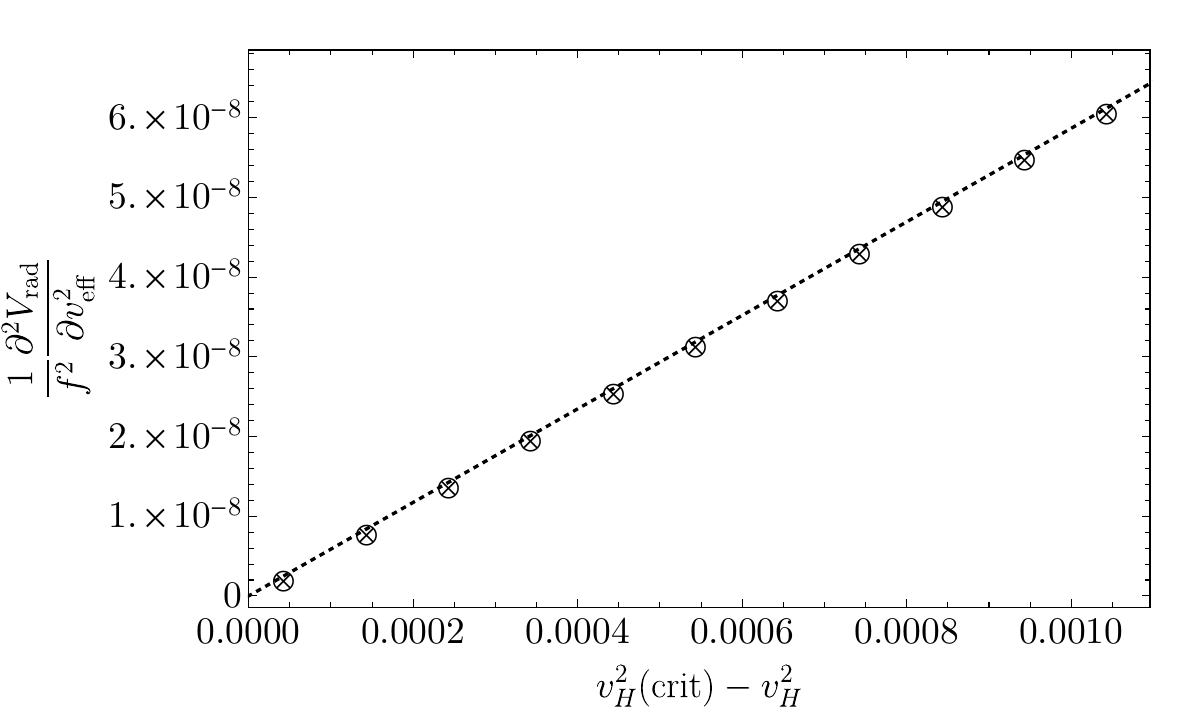}\includegraphics[width=0.5\textwidth]{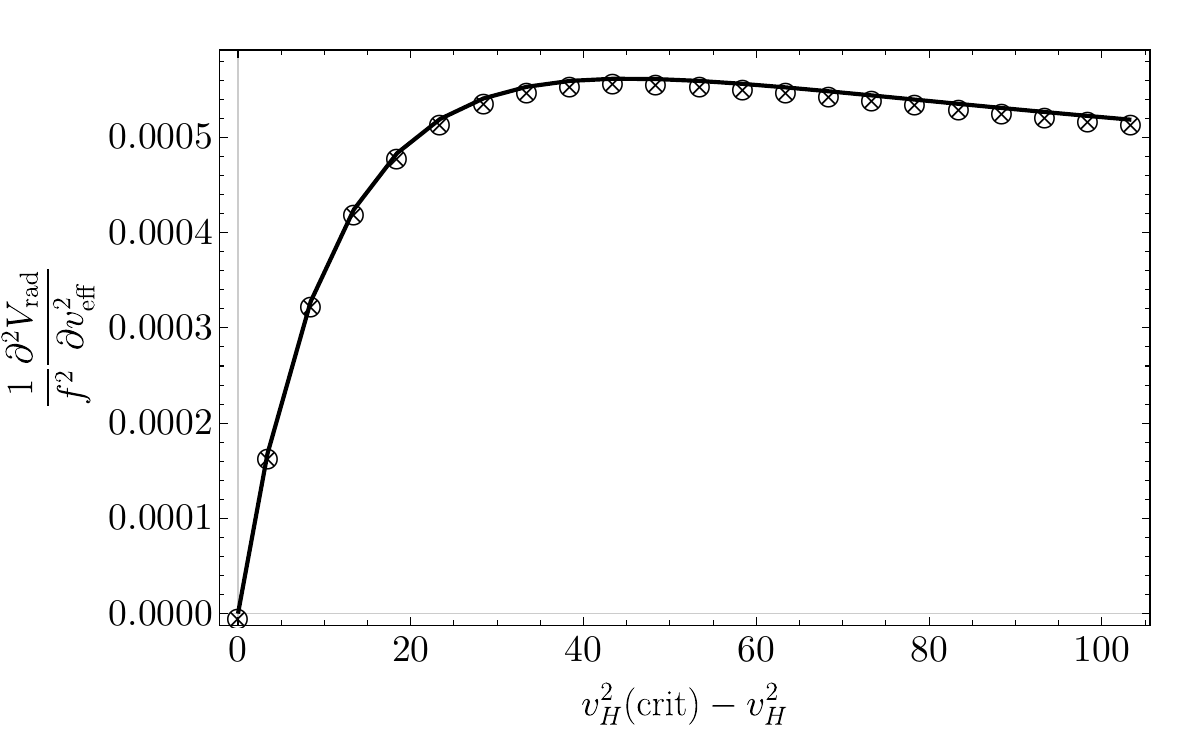}
	\caption{\small In this Figure, we display the curvature of the radion potential as a function of the effective Higgs VEV for $v_H^2 < v_H^2 (\text{crit})$.  In the plot on the left, we focus on $v_H^2$ close to the critical value, while on the right, we display the curvature for a wider range of $v_H^2 < v_H^2 (\text{crit})$.  Near the critical $v_H^2$, the behavior is well described by a line intersecting with the origin, with the critical value here determined to be $v_H^2(\text{crit}) = -16.58305645$, apparently consistent with the value determined on the sub-critical side in Figure~\ref{fig:vhcrit} up to numerical errors in the solving routine.}
	\label{fig:vhcritsup}
\end{figure}

In Figure~\ref{fig:vhcritsup}, we display the behavior of the model with supercritical values of $v_H^2$, focusing on the value of the second derivative of the radion potential at the minimum.  The second derivative remains positive for all $v_H^2$ values less than the critical one, and so the minimum of the potential along the line where the Higgs boundary conditions are met coincides with the Higgs critical point.  

While the radion potential has a smooth minimum (as a function of the effective Higgs VEV) in this dynamical model on either side of the critical value of $v_H^2$, we again find that it does not generally satisfy the $\mu\nu$ components of Einstein's equations, specifically the metric junction conditions $\sqrt{G_{0,1}}= \pm \frac{\kappa^2}{6} V_{0,1}$.  


Of course one could fine-tune both tensions to meet both metric junction conditions, as is done in the original unstabilized Randall-Sundrum model, however one should first ask what physical phenomena occur when such tuning is not performed.  In the RS model, the mistunes lead to either collapse or runaway of the branes, however in this case, it seems unlikely that the same behavior occurs, as the potential does not exhibit obvious runaway directions.  What seems more likely is that the background solution obtained after relaxing the metric ansatz to include bent branes (e.g. nontrivial time dynamics).  This is discussed further in Section~\ref{sec:cosmo}.

\section{CFT Interpretation}
\setcounter{equation}{0}
\label{sec:CFTInterpretation}

Here we comment on the 4-dimensional CFT interpretation of this model.  The dual of this picture has a parallel in weakly coupled models where electroweak symmetry breaking is driven radiatively, as is the case in the MSSM~\cite{Ibanez:1982fr}.  In such cases, the electroweak scale arises via dimensional transmutation, with renormalization group effects creating the instability that is rectified by the vacuum expectation value of the Higgs in spite of the microscopic theory having no explicit scales.

In the picture under consideration, a similar instability is reached when the scaling dimensions of operators in a quasi-conformal theory are pushed towards and potentially into the complex plane through renormalization group flow.  Such complex scaling dimensions are a usual part of the description of theories with a discrete scale invariance, thus we have a picture where a theory evolves off of a standard nontrivial UV fixed point and begins to exhibit discrete scale invariance in the IR.  Discrete scale invariance is found in the IR behavior of the Higgs profile, where the Higgs behaves approximately as
\beq
\phi \propto z^{2-\epsilon/4} \cos \left( \sqrt{\lambda} \log z + \gamma \right).
\eeq
The solution is simple scaling under the discrete transformation $z \rightarrow z \exp\left(2 \pi/\sqrt{\lambda}\right)$, corresponding to a discrete scale transformation $\mu \rightarrow \mu \exp\left(-2 \pi/\sqrt{\lambda}\right)$.  While at first glance interesting, discrete scaling behavior is forbidden in the deep IR~\cite{Luty:2012ww},  and is expected to be terminated in some way -- likely by the formation of condensates and a transition scale past which RG flow resumes more standard behavior.  Indeed, the study of the scalar fluctuations in Section~\ref{sec:runningmass} shows that if one tries to continue the bulk too far into the regime of log-periodic behavior, additional tachyons emerge that are, at least in the toy 5D theory of Section~\ref{sec:runningmass}, unresolved.

In the dynamical model of Section~\ref{sec:gwdriven}, the driving scalar field $\phi_\textnormal{d}$ plays the role dual to an operator whose coupling runs slowly, slightly deforming the CFT, with the deformation growing in the infrared.  This running backreacts, in general, on the theory, and can generate a running for scaling dimensions of other operators in the theory.  The bulk trilinear coupling between the Higgs and the scalar $\phi_\textnormal{d}$ is the pathway for this backreaction.  

The running scaling dimension of the operator associated with the bulk Higgs is dual to the 5D $z$-dependence of the effective 5D mass of the Higgs in the background of the scalar, $\phi_\textnormal{d}$, and the instability associated with complex scaling dimensions and discrete scale invariance is dual to supersaturation of the Breitenlohner-Freedman bound.  Since the effective mass begins above the BF bound, the far UV behavior is that of a normal CFT without instabilities, where the operators have normal real scaling dimensions.  It is only in the IR behavior, where scaling dimensions become complex, that an instability emerges.

This class of instability, without the dynamical aspect we explore, was investigated in work by Kaplan, Lee, Son, and Stephanov~\cite{Kaplan:2009kr} along with a proposed AdS dual.   In this work, it was conjectured that the loss of conformality as a function of some descriptor of the theory, such as the number of massless QCD flavors, can be thought of as due to the annihilation of two fixed points (UV and IR) under variations of that parameter.   They posit that such a theory could contain some operator $O$ that has different scaling dimensions at the two fixed points, $\Delta_{\text{UV, IR}}$, and these scaling dimensions smoothly merge and become complex when the external descriptor is moved past some critical value.  It was pointed out in this work that the behavior of the theory under this transition is closely similar to scaling behavior associated with finite temperature topological phase transitions of the sort studied by Berezinskii, Kosterlitz, and Thouless (BKT)~\cite{Berezinsky:1970fr,Kosterlitz:1973xp}.  In these models, there is a critical line, with the theory being conformal for a finite range of descriptor, and with a gap turning on smoothly past the point where conformality is lost.  These two scaling dimensions, $\Delta_\text{UV, IR}$, correspond in the 5D AdS dual to the two solutions to the scalar equations of motion in the $z \rightarrow 0$ region, each of which has different scaling properties given by $\Delta_\text{UV}$ and $\Delta_\text{IR}$.  In 5D, the loss of conformality corresponds to the merging of these two scaling solutions at the Breitenlohner-Freedman bound as the bulk mass is taken through $m^2 = -4$.  Below the BF bound, the theory requires a UV cutoff, and also predicts an IR scale associated with rectification of a tachyon instability through condensation of bulk fields, corresponding in the holographic picture to a VEV for the operator $O$.

\begin{figure}[t]
	\centering
	\begin{subfigure}[t]{0.49\textwidth}
		\includegraphics[height=1.75in]{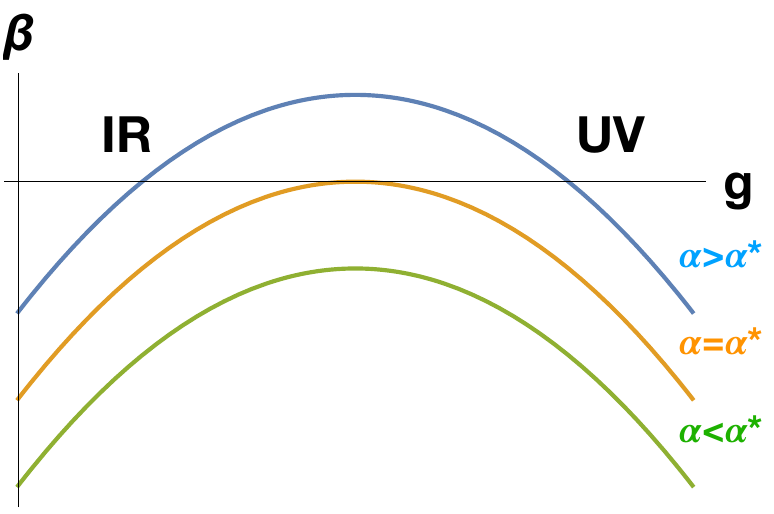}
	\end{subfigure}~
	\begin{subfigure}[t]{0.49\textwidth}
		\includegraphics[height=1.75in]{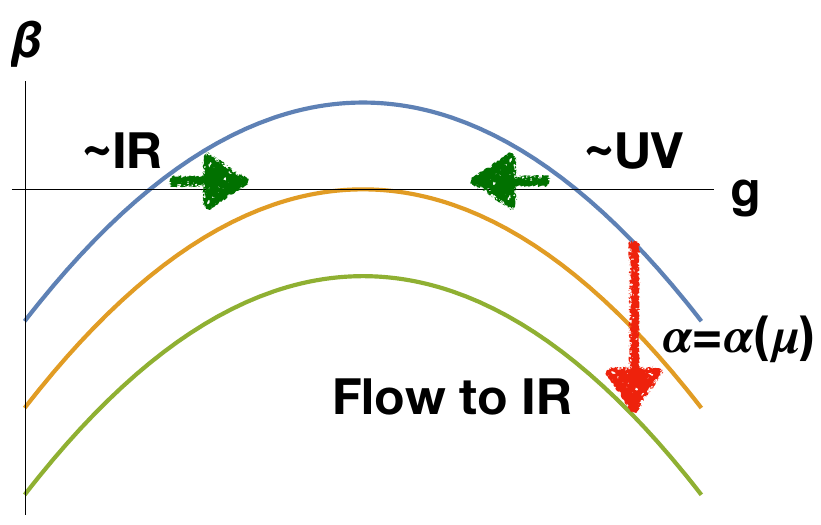}
	\end{subfigure}
	\caption{\small Here we show a cartoon of an approximate CFT dual of our 5D model.  On the left is the picture of fixed points annihilating under continuous variation of some descriptor of the theory, as explored in~\cite{Kaplan:2009kr}.  On the right is our picture of quasi-fixed points annihilating under renormalization group evolution.}
	\label{fig:dynamicalconflost}
\end{figure}

The 5D model we have described has given dynamics to this picture, where what was an external parameter has been promoted to a coupling in the theory which has nontrivial RG evolution.  In Figure~\ref{fig:dynamicalconflost}, we give a cartoon of what the model we explore achieves.  In~\cite{Kaplan:2009kr}, in the case that parameters are chosen to put the theory in a conformal window, there are explicit UV and IR fixed points, both nontrivial.  Moving in and out of the conformal window is achieved by varying those external parameters, with the fixed points merging at its threshold.  In the case of our model, the idea is that the theory begins at or flows quickly to an IR fixed point which has been demoted by a slightly relevant deformation of the theory to a quasi-fixed point.  The theory tracks this IR quasi-fixed point until it disappears after annihilating its associated UV quasi-fixed point.  Under further RG flow, scaling dimensions become complex with a corresponding discrete scaling law, the theory becomes unstable, and the instability is potentially resolved by condensates.  The theory can also begin and remain near the UV quasi-fixed point, in principle, corresponding to taking the tuned boundary condition for the bulk scalar that picks out the other slower-growing solution.

When the instability is rectified by condensates,  the approximate conformal invariance is broken spontaneously.  There are different options for this breaking.  The Higgs itself can form a vacuum expectation value, likely along with condensates of the operator that is driving the theory towards the instability.  This option gives a Higgs mass and 4D effective VEV that is not typically suppressed in comparison with the 5D KK scale or its dual picture compositeness scale.  There are other options, however.  It is known that the phase structure of superconductivity can be quite rich, allowing for condensates with inhomogeneous spatial configurations such as stripes or crystalline structure, and this has been reproduced in the holographic context~\cite{Hartnoll:2016apf}.  

A dynamical dilaton field, corresponding to the condensates of operators in the CFT and fluctuations about these points in field space, has a potential that depends on which operators take on VEVs, and its potential selects the most attractive channel for resolving the tachyon.  The extra-dimensional modulus field, the radion, is the dual to this dilaton.  Finding the classical extrema of the 5D effective potential corresponds to identifying the vacuum state of the dual CFT.

When sourced operators with nontrivial scaling dimension take on vacuum expectation values, approximate conformal invariance is spontaneously broken, and the dilaton potential is nontrivial (more than a scale-invariant quartic).  If the potential has a nontrivial minimum, the resulting mass of the dilaton particle is proportional to the degree of explicit breaking of conformal symmetry.  In our picture, as a function of the gap of the CFT associated with a VEV of a marginally relevant operator $O_\epsilon$, an IR tachyon instability eventually emerges for smaller values of $\langle O_\epsilon \rangle$, in the regime of discrete scale invariance.   At this point the VEV $\langle O_H \rangle$ turns on in addition to $\langle O_\epsilon \rangle$.

The kink in the toy model of Section~\ref{sec:runningmass}, or its cousin, the non-existence of solutions with nonvanishing Higgs VEV past some critical $z_1$ in the dynamical model of Section~\ref{sec:gwdriven}, are the most curious features of this setup.  These behaviors appear to be strongly tied to the type of instability that is dominant in the model.  The tachyon that emerges due to the renormalization group instability/violation of the BF bound appears to be a necessary component for novel behavior of the potential.  This hypothesis is supported by what is observed in Figure~\ref{fig:v2plots}, and in the behavior of the scalar solutions in Section~\ref{sec:gwdriven}.  Taking, for example, $v_H^2$ to be large, giving a large 4D tachyonic mass to the Higgs on the brane, gives a more standard kind of picture where a Higgs condensate turns on and determines the brane location.  However, for smaller or negative $v_H^2$, when there is no longer a 4D brane-localized tachyon, there is a turnaround in the behavior of the solutions.  For $v_H^2$ below the critical value, it is the bulk mass falling below the BF bound that is the more important component of the instability.  

\section{Discussion}
\setcounter{equation}{0}
\label{sec:discussion}

\subsection{Connections to Condensed Matter and Statistical Physics}
The construction we explore potentially gives a new way to think about self-organized criticality in the fields where it was first explored and named~\cite{soc}.  The scaling of perturbations ($1/f$ flicker noise in the literature), and the instabilities associated with catastrophic failure and its possible connection to an emergent discrete scale invariance have their signatures in the purported holographic dual we have proposed.
The $1/f$ so-called ``flicker" noise (in fact $1/f^\alpha$, where $\alpha$ can depend on the system under consideration) is simply the signature of criticality itself~\cite{socflicker} -- a fixed point where perturbations exhibit scaling laws.  There should be CFTs describing the coarse-grained effective theory of such systems where criticality is self-organized, and there could be AdS dual descriptions of such theories.  The scaling laws associated with the self-organized critical point are associated in this case to the scaling behavior of field solutions in AdS space.
The above discussion applies to any critical point, not just self-organized ones.  However, the particular 5D picture we have presented here appears to have features which place it close to systems with self-organization.  These share the commonality that the systems in question are brought to the point of some sort of failure mode that leads to time dynamics, typically a form of adjustment.  It has been suggested that the development of the adjustment behavior can be associated, in the coarse-grained theory, with scaling dimensions of the critical point being driven into the complex plane, and creating a discrete scale invariance, with fluctuations obeying a log-periodic power law.  In the context of quantum field theory, it is known that such scaling laws are not allowed, and have instabilities associated with them~\cite{Luty:2012ww}.  In our picture these instabilities map to the region in the radion potential in which there are unresolved tachyons.  Should the system be placed at these points of instability, it is, at the moment, unknown what the response of the system will be.

\subsection{Incorporation of the Standard Model}

The eventual goal is to embed this class of Higgs sector into a theory which accommodates the rest of the SM field content, and where the Higgs resides not precisely at the point of criticality, but instead picks a VEV and breaks the electroweak gauge symmetry spontaneously.  This may occur as a result of finite radiative corrections, or perhaps through nontrivial feedback due to explicit breaking of conformal invariance in the SM (for example, from confinement and chiral symmetry breaking in QCD), or in extensions of it (as in~\cite{Graham:2015cka}).  If the SM can accommodate such a Higgs sector, a key calculation will be the Higgs cubic coupling in the context of the manner in which its potential is generated, as this will be eventually probed by colliders.  As we will discuss in the next section, cosmology of such an extension of the SM could be very interesting.  As we have emphasized, we have not yet fully identified the vacuum state, which we argue must be time-dependent unless tuning is performed.  

\section{Speculation: Cosmology}
\setcounter{equation}{0}
\label{sec:cosmo}
The phenomenology of this scenario, if employed by nature in creating a low electroweak scale, is expected to be quite novel.  Cosmology stands out as a particularly interesting area, due to the interplay between the radion and the Higgs.  Metastability of the self-organized critical state is a vital consideration, although possibly resolved trivially by details of the deformation that drives self-organization or by a more complete model without a hard wall.  A very interesting possibility is that dynamics of the radion could ``rock" the early universe across the electroweak phase transition, sourcing a stochastic gravitational wave background, and creating an era of the early universe with an exotic equation of state (leading to modified constraints on inflationary scenarios and/or moduli masses)~\cite{Amin:2018kkg}.  This may be interesting also from the standpoint of baryo- and leptogenesis.  

Even more curious features are likely to emerge under a full calculation of the classical background, which we have left for future work.  As we have emphasized, in writing down a 4D Lorentz-invariant radion potential, we have been required to do some violence to the theory.  That is, we have calculated the potential under the presumption that the metric slices are flat.  However, at the minima we have identified, the consistency conditions for a flat metric ansatz on the boundaries of the space, that both the UV and IR contributions to the radion potential separately vanish at the minimum, cannot be met without tuning.  Bent 4D slices thus appear to be an integral part of the full solution to the theory at or near Higgs criticality.  The theory is telling us that, as part of the resolution of the AdS tachyon, it breaks Lorentz invariance spontaneously at long distances. It is interesting to speculate at the form that this takes without (at least in this work) undertaking a detailed calculation.

Pessimistically, we might imagine that the solution could be a runaway.  This seems unlikely from the following consideration of the dynamical model discussed in Section~\ref{sec:gwdriven}:  one could imagine adjusting the bulk cubic coupling $\lambda$ between $\phi_\textnormal{d}$ and the Higgs, increasing it starting from some small value.    The system would then reside at the minimum of the Goldberger-Wise potential created by the VEV of the $\phi_\textnormal{d}$ scalar, and the Higgs would have positive mass squared.  One would in this case start in an unbroken phase with a massive Higgs field near the KK scale, and a completely static geometry (presuming the usual single tuning of  the UV brane tension).  Increasing the coupling would move the turn-on of the Higgs VEV ever closer to the minimum of the Goldberger-Wise potential, the critical $z_1$ value approaching the minimum from large $z_1$.  The lowest-lying Higgs excitation would then gradually move to the bottom of the spectrum.  There is a fine-tuned value of $\lambda$ where the Higgs extremum coincides exactly at the minimum of the GW potential.   The theory at this point, as the usual solution to the theory at the Goldberger-Wise minimum tells us, has a massive radion.  Due to the adjustment of $\lambda$, it also has a (finely-tuned) massless Higgs.  The only nearby instability in the low energy theory seems like the usual one associated with the Higgs phase transition.  Further increase in $\lambda$ either results in a Higgs VEV and symmetry breaking of the usual sort, or a more novel transition.   It would appear the latter is a good possibility in some fraction of the parameter space of the model.

We have shown that the change in the radion potential due to the Higgs contribution creates a new minimum and for large negative $v_H^2$, there appears to be an onset of a novel type of transition that does not satisfy the usual metric junction condition.  This means that the transition is not likely  a normal Higgs one, but rather could be a transition to a spontaneously Lorentz-violating dynamical background.  The radion should still be stabilized by its mass if we remain close to the boundary of the critical region.  Therefore if it moves, it likely oscillates rather than runs away, acting ``trapped"~\cite{Kofman:2004yc}.\footnote{We thank Ofri Telem for suggesting the possibility of an oscillating radion background.}  Amusingly, such a ``trapped" modulus, the oscillating radion field, might look cosmologically similar to non-relativistic matter from the standpoint of the low energy theory: a Bose-Einstein condensate of radion particles.  Deeper into the critical Higgs region, as the metric junction conditions become further from being satisfied, this time dependence may become very complicated, involving a mixture of the many degrees of freedom in the model.  A simulation including full backreaction may be needed.  It will be interesting to see more clearly what the low energy theory looks like with further study.

\begin{figure}[t]
	\centering
	\includegraphics[height=2.5in]{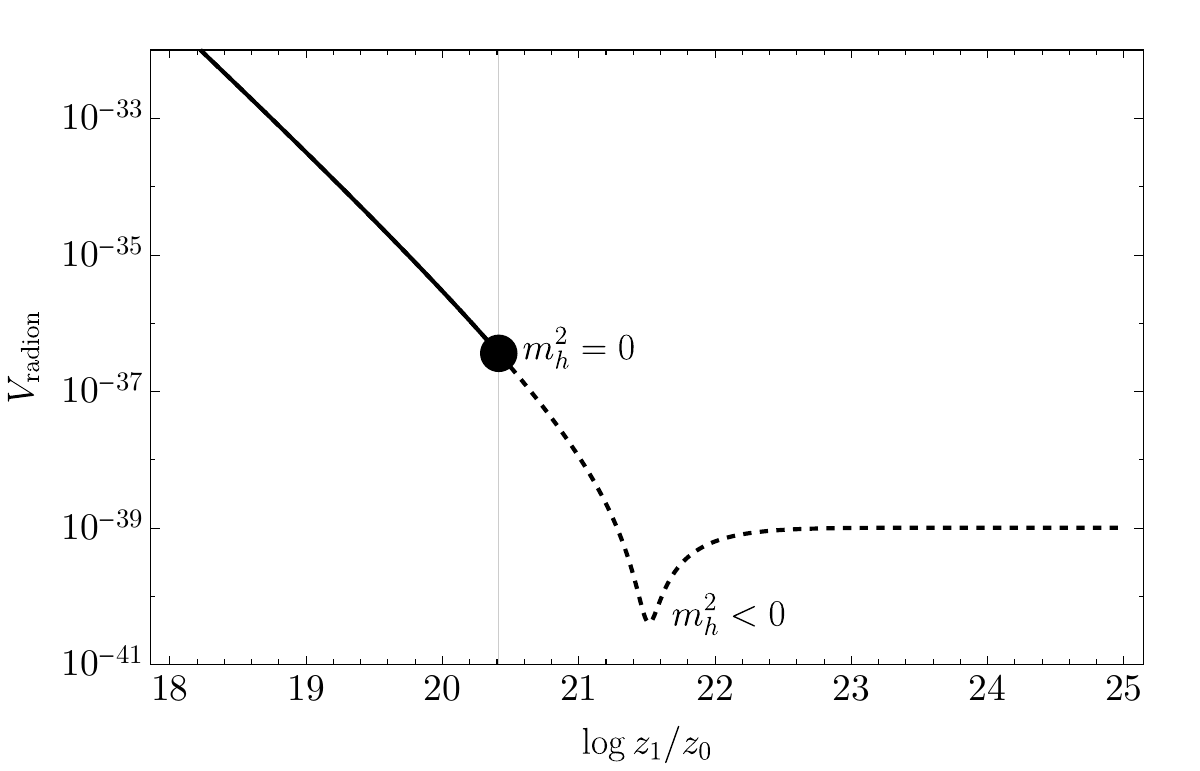}
	\caption{Here we show the radion potential. The dashed line is the potential if the Higgs VEV is left vanishing.  There is a minimum of this potential where the metric ansatz for the IR brane is satisfied, but it corresponds to an unstable Higgs configuration.  The solid line is the region where the effective Higgs mass squared in the low energy theory is positive.  At the dot, the mass vanishes, and if $v_H^2 < v_H^2(\text{crit})$, the potential of the radion is minimized if the unstable Higgs region is forbidden.  The gravity sector is not extremized here -- the metric junction condition on the IR brane is not met.}
	\label{fig:frustrated}
\end{figure}

In short, the resolution of the IR-emergent AdS tachyon appears to require a spontaneous breakdown of 4D Lorentz invariance possibly a time-dependent  oscillatory vacuum state.  In the language of condensed matter physics and superconductivity literature, the theory would be resolving the instability by entering a ``striped phase" at long distances with oscillations of the radion state breaking time translation invariance.  The generic phenomenon of classical spontaneous breakdown of time translation invariance was investigated in~\cite{Shapere:2012nq}, with a quantum mechanical version explored in~\cite{Wilczek:2012jt}.  These striped phases in condensed matter systems appear to arise in the presence of a frustration in the system where competing phenomena strive to set the vacuum state.  In the case of our radion model, there are two competing ``minima" of the potential: one where the radion is at a global minimum set by the Goldberger-Wise mechanism, but the Higgs is unstable, and the other where the Higgs field is stabilized, but the 5D gravity sector is not at an extremum of the action.  The outcome in some condensed matter systems is to resolve such tension by entering a translation invariance-breaking striped phase, and we are suggesting similar physics may be at work in resolution of this AdS tachyon, although the breaking could be of time-translation invariance.  A picture of these competing extrema is shown in Figure~\ref{fig:frustrated}.

This sort of time dependent vacuum state and its application to cosmology was first explored in~\cite{Bains:2015gpv}, with important recent followup work concerning stability of this scenario in~\cite{Easson:2018qgr}.  

In the 4D CFT language, there may be strongly coupled quasi-conformal 4D theories with a dynamically generated gap that have aspects of matter-$\Lambda$ cosmology built into their ground state at long distances due to frustration of the dilaton.  Classical breaking of time-translation invariance in the 5D theory corresponds to its quantum mechanical counterpart in the 4D dual.  The apparent puzzle of the light and seemingly unprotected Higgs in the critical region may be that it is the Goldstone boson of this breaking, a type of phonon, perhaps making the lightness of the Higgs directly connected to the presence of a cosmology with non-relativistic matter and dark energy content.  

A better understanding of the dynamical aspects of this class of Higgs model is definitely required to make these statements more firm, but there seem to be some quite promising avenues to pursue.


\section{Conclusions}
\setcounter{equation}{0}
We have discussed a new possible approach to the Higgs hierarchy problem.  The model is in part inspired by attempts to model aspects of self-organized criticality in condensed matter systems in which it has been hypothesized that some classes of these critical states on the brink of catastrophic failure contain critical exponents that are becoming complex under loading of the system, leading to discrete scale invariance and instabilities.  

We have explored 5D constructions that have features that are similar to the behavior just described.  It has made us directly confront the AdS tachyon, associated with violation of the Breitenlohner-Freedman bound, and search for the manner in which field theory might resolve it.  In this model, the resolution takes the form of an IR brane with characteristics that depend on the 5D fundamental parameters.  We found that there is are large regions of the parameter space where a novel type of transition appears to be taking place.  

%

A dynamical cosmology is an unavoidable consequence of the model when placed inside or near this region.  Time evolution appears crucial to resolution of the AdS tachyon in this model.  This could be a feature rather than a bug, tying together puzzling aspects of fundamental particle physics with puzzling features of cosmology in a novel way.  It remains to be seen what the Higgs vev and spectrum will be once this time evolution is completely taken into account, and whether its effective mass and/or vacuum expectation value are tied in some interesting way to aspects of the cosmology.

Top priorities are to further investigate the cosmological dynamics, determine whether this type of setup can be utilized as an extension of the Standard Model, study its novel phenomenological implications and constraints if so, and to seek an embedding in a more complete microscopic theory.

\section*{Acknowledgments}

The authors thank Simon Catterall, Csaba Cs\'aki, Michael Geller, Yuval Grossman, Sungwoo Hong, Marek Olechowski, Stefan Pokorski, Prashant Saraswat, and John Terning for helpful discussions at various stages of this project.  The authors also thank Csaba Cs\'aki, Michael Geller, and Ofri Telem for helpful comments on this draft in the later stages of preparation.  The authors thank Cornell University for hospitality throughout the duration of this project.  JH thanks the Aspen Center for Physics for hospitality during part of this research project.  The ACP is supported by National Science Foundation grant PHY-1607611.  JH also thanks the University of Grenada and the University of Warsaw for hospitality during extended visits.  CE, JH, and GR are supported in part by the DOE under grant award number DE-SC0009998.

\appendix
\section{Removal of the Kink in Dynamical Models}
\setcounter{equation}{0}
\label{app:kink}
The radion potential in the dynamical models studied is smooth at the critical point, rather than kinked, as in the toy model of Section~\ref{sec:runningmass}.  There are many contributions to the potential, some of which arise from backreaction of the Higgs VEV onto the mechanism which is driving the effective bulk Higgs mass, the $\phi_\textnormal{d}$ scalar in this case.  Such contributions are absent in the model with an explicitly varying mass term, and in the dynamical model, they remove the kink behavior of the radion potential.  In order to see that the potential is smooth, we start from the radion potential in the small backreaction limit of Eq.~\eqref{eq:radpotGW}. Moreover, for the sake of concreteness, we consider arbitrary brane potentials $V_{0,1}^H$ for the Higgs, but we focus on the simple model of Section~\ref{sec:gwdriven} in which a stiff wall boundary condition fixes $\phi_\textnormal{d}$ to $v_0$ on the UV brane, and the IR brane potential for $\phi_\textnormal{d}$ consists only of a brane-localized mass term, $V_1^{\phi_\textnormal{d}}=-\epsilon\phi_\textnormal{d}^2$. The proof is analogous for other brane potentials, for example when taking a stiff wall boundary condition on both branes for $\phi_\textnormal{d}$, although more computationally involved.

It is sufficient for us to show that the derivative of the radion potential does not change when going across the critical point $z_\textnormal{c}$, i.e.\ it has the same value at the critical point with and without the Higgs VEV turned on. Since $h_0^2$ is approximately linear in $z_1-z_\textnormal{c}$, this is equivalent to showing that the terms in the radion potential that are quadratic in $h_0$ cancel against each other at the critical point. So, using the parametrization $\phi_h = h_0 f_h(z)$, let us consider the radion potential to quadratic order:
\beq
\begin{aligned}
	V_\textnormal{rad}^{(2)}&=z_0^{-4}\left[V_0^{(2)}-\frac{1}{2}\epsilon v_0z_0^{4-\epsilon}I_4-\frac{1}{4}z_0^2h_0^2f_{h,0}'^2+\frac{1}{4}\left(m_H^2-\lambda v_0\right)h_0^2f_{h,0}^2\right] \\
	&\phantom{{}={}}+z_1^{-4}\left[V_1^{(2)}+\frac{1}{4}z_1^2h_0^2f_{h,1}'^2-\frac{1}{4}\left(m_H^2-\lambda v_0\left(\frac{z_1}{z_0}\right)^{\!\epsilon}\,\right)h_0^2f_{h,1}^2\right].
\end{aligned}
\eeq
Now, the key point is to realize that, at the critical point, the only term involving the integral reduces to
\beq
\begin{aligned}
	\epsilon v_0z_0^{-\epsilon}I_4&=\frac{\lambda h_0^2}{2}\epsilon v_0z_0^{-\epsilon}\int_{z_0}^{z_1}f_h^2\xi^{-5+\epsilon}d\xi \\
	&=-\frac{h_0^2}{2}\int_{z_0}^{z_1}f_h^2\xi^{-4}\left(m_H^2-\lambda\phi_\textnormal{d}\right)'d\xi \\
	&=\frac{h_0^2}{2}z_0^{-4}\left[-z_0^2f_{h,0}'^2+4z_0f_{h,0}f_{h,0}'+\left(m_H^2-\lambda v_0\right)f_{h,0}^2\right] \\
	&\phantom{{}={}}+\frac{h_0^2}{2}z_1^{-4}\left[z_1^2f_{h,1}'^2-4z_1f_{h,1}f_{h,1}'-\left(m_H^2-\lambda v_0\left(\frac{z_1}{z_0}\right)^{\!\epsilon}\,\right)f_{h,1}^2\right],
\end{aligned}
\eeq
where we have integrated by parts and used the equation of motion for $f_h$, so that it was possible to perform the residual integral explicitly. At this point, the quadratic parts of the brane potentials can be written in the following way by imposing the Higgs boundary conditions:
\beq
V_0^{(2)}=z_0h_0^2f_{h,0}f_{h,0}', \qquad \qquad V_1^{(2)}=-z_1h_0^2f_{h,1}f_{h,1}'.
\eeq 
So, after plugging in the expression for the integral, one finds that the quadratic Higgs contributions to the radion potential vanish at the critical point.  Thus, there is no kink in the scalar-driven model with these boundary conditions.  A proof of smoothness for the stiff wall boundary conditions $\phi_\textnormal{d}(z_{0,1}) = v_{0,1}$ proceeds similarly, with the same final result.



\end{document}